\documentclass[amsmath, amssymb, amsfonts, 12pt, nofootinbib, a4paper, prd, notitlepage, preprint, a4paper]{revtex4-2}

\usepackage{hyperref}
\usepackage{graphicx}
\usepackage{bm}\usepackage{slashed}
\usepackage{dsfont}
\usepackage{braket}
\usepackage{xcolor}

\newcommand{\lgr}{\mathcal{L}}
\newcommand{\M}{\mathcal{M}}
\newcommand{\D}{\textnormal{d}}

\newcommand{\dmu}{\partial_{\mu}}
\newcommand{\dnu}{\partial_{\nu}}

\newcommand{\Mt}{\tilde{M}_{\rm{Pl}}}
\newcommand{\Mp}{\tilde{M}'}
\mathcode`*=\string"8000
\begingroup
\catcode`*=\active
\xdef*{\noexpand\textup{\string*}}
\endgroup

\usepackage{setspace}

\begin{document}


\title{Fifth forces and broken scale symmetries in the Jordan frame}

\author{Edmund J.\ Copeland}
\email{ed.copeland@nottingham.ac.uk}
\affiliation{School of Physics and Astronomy, University of Nottingham,\\
Nottingham NG7 2RD, United Kingdom}

\author{Peter Millington}
\email{p.millington@nottingham.ac.uk}
\affiliation{School of Physics and Astronomy, University of Nottingham,\\
Nottingham NG7 2RD, United Kingdom}

\author{Sergio Sevillano Mu\~{n}oz}
\email{s.sevillano@nottingham.ac.uk}
\affiliation{School of Physics and Astronomy, University of Nottingham,\\
Nottingham NG7 2RD, United Kingdom}

\date{16 February 2022}

\begin{abstract}
    We study the origin of fifth forces in scalar-tensor theories of gravity in the so-called Jordan frame, where the modifications to the gravitational sector are manifest. We focus on theories of Brans-Dicke type in which an additional scalar field is coupled directly to the Ricci scalar of General Relativity. We describe how the necessary diffeomorphism invariance of the modified gravitational sector leads to a modification of the usual gauge fixing term (for the harmonic gauge), as compared to Einstein gravity.  This allows us to perform a consistent linearization of the gravitational sector in the weak-field limit, which gives rise to a kinetic mixing between the non-minimally coupled scalar field and the graviton. It is through this mixing that a fifth force can arise between matter fields. We are then able to compute the matrix elements for fifth-force exchanges directly in the Jordan frame, without the need to perform a conformal transformation to the so-called Einstein frame, wherein the gravitational sector is of Einstein-Hilbert form. We obtain results that are in agreement with the equivalent Einstein-frame calculations and illustrate, still in the Jordan frame, the pivotal role that sources of explicit scale symmetry breaking in the matter sector play in admitting fifth-force couplings. \\~\\
    \begin{footnotesize}This is an author-prepared post-print of \href{https://doi.org/10.1088/1475-7516/2022/02/016}{JCAP02(2022)016}, published by IOP Publishing Ltd.\ under the terms of the \href{https://creativecommons.org/licenses/by/4.0/}{CC BY 4.0 license}.\end{footnotesize}
\end{abstract}


\maketitle


\newpage

\section{Introduction}
    
    Without symmetries to prevent it, scalar fields that are realised in nature will inevitably couple to any sector of a particle physics model that knows about gravity.  This might, e.g., be through Higgs portals \cite{Silveira:1985rk, McDonald:1993ex, Burgess:2000yq, Davoudiasl:2004be, Schabinger:2005ei,Patt:2006fw}, through neutrino portals \cite{Falkowski:2009yz,GonzalezMacias:2015rxl} or through direct couplings to curvature invariants, such as the Ricci scalar in the case of theories of Brans-Dicke type~\cite{Brans:1961sx}. However, the ability to make Weyl rescalings in metric theories of gravity makes it difficult to disentangle these various ways that scalar fields can couple to matter~\cite{Burrage:2018dvt, Millington:2019jbx}. Theories of gravity in which scalar fields couple directly to curvature terms are often referred to as scalar-tensor theories \cite{FM}. 
    
    In fact, as far as quantum effects are concerned, any theory of gravity plus a scalar field should be regarded as a scalar-tensor theory: If, e.g., we tune away a direct coupling to the Ricci scalar at tree level or at a given energy scale, it will be regenerated by loop corrections or the renormalization group running to another scale (see, e.g., Refs.~\cite{Herranen:2014cua, Markkanen:2018bfx}). Hence, if we want to treat quantum field theories that know about gravity and involve scalar fields, we should treat them in the so-called Jordan frame (or Jordan-like frames), wherein all of the allowed couplings are present, be they to the gravitational or non-gravitational sectors. The import is that the so-called screening mechanisms (see, e.g.,~\cite{Joyce:2014kja, Burrage:2017qrf}) that can allow scalar-tensor theories to evade local tests of gravity can similarly apply to Higgs-portal theories~\cite{Burrage:2018dvt, Brax:2021rwk}.  We will not take screening into account in this work, but examples include the chameleon~\cite{Khoury:2003aq, Khoury:2003rn}, symmetron~\cite{Hinterbichler:2010es, Hinterbichler:2011ca} and Vainshtein~\cite{Vainshtein:1972sx} mechanisms, and their variants. For recent reviews on experimental and observational constraints (on screened) fifth forces, see Refs.~\cite{AlvesBatista:2021gzc,Burrage:2017qrf,Brax:2021wcv}.
    
    One symmetry that can affect the way that additional scalar degrees of freedom from the gravitational sector couple to matter is scale symmetry (or, in the local case, Weyl symmetry). In general, these scalars will mediate fifth forces between matter sources, and new long-range forces of nature are heavily constrained \cite{Burrage:2017qrf}. However, if the matter sector is scale invariant, the classical fifth forces are suppressed. In the case of the Standard Model (SM) of particle physics, the scale breaking term that plays a dominant role in allowing gravitational scalars to couple to matter is the quadratic term of the SM Higgs potential. This becomes clear when one works in the Einstein frame, as was described in Ref.~\cite{Burrage:2018dvt}. On the other hand, the role played by the same operator is less obvious when considered in the Jordan frame.
    
    A popular example of a model that exploits scale symmetry to prevent the emergence of fifth forces is the Higgs-dilaton theory~\cite{Wetterich:1987fm, Buchmuller:1988cj, Shaposhnikov:2008xb, Shaposhnikov:2008xi, Blas:2011ac, Garcia-Bellido:2011kqb, Garcia-Bellido:2012npk, Bezrukov:2012hx, Henz:2013oxa, Rubio:2014wta, Karananas:2016grc, Ferreira:2016vsc, Ferreira:2016kxi, Casas:2017wjh, Ferreira:2018qss}. Therein, both the Higgs field and an additional gauge-singlet scalar are non-minimally coupled to the Ricci scalar. The action has a global Weyl symmetry, which is broken when the fields acquire vacuum expectation values (vevs) along a particular trajectory in field space. Note that the structure of the Higgs potential remains to be probed experimentally~\cite{Maltoni:2018ttu,Borowka:2018pxx,ATLAS:2019pbo,CMS:2018ipl,Tian:2016qlk,Abramowicz:2017,Agrawal:2019bpm,Nielsen:2020uvd}. However, due to the dynamical origin of the scale breaking, the massless mode does not couple to the matter sector, leaving a theory devoid of fifth forces~\cite{Ferreira:2016kxi,Brax:2014baa, Ferreira:2016vsc}. The role of explicit versus spontaneous (in this dynamical sense) scale breaking in the emergence of fifth forces was described in Ref.~\cite{Burrage:2018dvt}, where the analysis was performed in the Einstein frame. In this paper, we will show explicitly how the same arguments are borne out directly in the Jordan frame.
    
    The merits of dealing with scalar-tensor theories in the Jordan frame are threefold: First, there are theories for which an Einstein frame does not, in general, exist, such as Horndeski \cite{Horndeski:1974wa, Kobayashi:2019hrl}, beyond Horndeski \cite{Traykova:2019oyx, Gleyzes:2013ooa} and DHOST \cite{Langlois:2015cwa, Langlois:2018dxi} theories. Second, as discussed above, in an interacting quantum field theory, the Einstein frame may exist only at a particular energy scale, with loop corrections and the renormalization of couplings regenerating Jordan-like frames. Third, the conformal transformation to the Einstein frame, and the subsequent field redefinitions needed to bring the theory as close to being canonically normalized as possible (notwithstanding any curvature of the field-space metric) must be performed on a model-by-model basis and may not be easily automated.
    
    In order to proceed directly in the Jordan frame, we will need to work with linearized gravity, which cannot be decoupled (as it can be by construction if we work in the Einstein frame) due to the presence of the non-minimal coupling. By this means, we are able to isolate the kinetic mixings between the graviton and non-minimally coupled scalar field that can give rise to fifth forces. Most importantly, we find that the non-minimal coupling, through its effects on the geodesics of the spacetime, necessitates an update to the covariant derivative. Focusing on the (would-be) harmonic gauge, we determine the fifth-force potential by means of the non-relativistic limit of the scalar-mediated scattering matrix elements. Moreover, our results agree with those obtained in the Einstein frame, as per Ref.~\cite{Burrage:2018dvt}. 
	
	This paper is intended to be both explicit and pedagogical. As such, we begin in Sec.~\ref{second} with a review of previous studies on the relation between scale-symmetry breaking and fifth forces. After properly identifying the symmetries of Brans-Dicke-type \cite{Brans:1961sx} scalar-tensor theories in Sec.~\ref{thirdS}, we define a so-called scalar-harmonic gauge, by updating the harmonic gauge from Einstein gravity, and perform the subsequent linearization of the gravitational sector. From there, we consider the fifth-force contributions to the M\o{}ller scattering in Sec.~\ref{fourth} and obtain the Yukawa potential from its non-relativistic limit, concentrating on systems where the scale symmetry is broken either explicitly or dynamically. Our conclusions are presented in Sec.~\ref{fifth}. Additional details are provided in the Appendix.

    \section{Fifth forces and scale symmetry in the Einstein frame}\label{second}
    
	In this section, we review the Einstein-frame description of fifth forces, emphasizing the key role of explicit sources of scale breaking in the matter sector in allowing fifth forces to couple to matter fields. What follows is based heavily on Ref.~\cite{Burrage:2018dvt}.
	
	In the Jordan frame, the equations of motion for the class of scalar-tensor theories on which we will focus can be derived from the following action: 
	\begin{equation}
		S=\int \D^4{x} \sqrt{-g} \left[\frac{1}{2}F(X)R - \frac{1}{2}Z(X)g^{\mu\nu}\partial_\mu X \partial_\nu X - U(X)\right] +  S_{\rm{m}}[ g_{\mu \nu},\{\psi\}],
	\end{equation}
	where the real scalar field $X$ couples non-minimally to gravity through the function $F(X)$ and evolves subject to the potential $U(X)$. $Z(X)$ allows for a non-canonical kinetic term. In addition, $R$ is the Ricci scalar, defined in terms of the Jordan-frame metric $g_{\mu\nu}$, and $S_{\rm{m}} = \int \D^4{x} \sqrt{-g} {\cal L}_m$ is the matter action, containing the set of matter fields $\{\psi\}$.  Throughout this article, we employ the mostly plus signature convention $(-,+,+,+)$. 
	
	We now have a choice:~we can either work directly in the Jordan frame, as we will do later in this article, or we can eliminate the direct coupling of the field $X$ to gravity, by making a Weyl rescaling of the metric, and work in the Einstein frame. Therein, the gravity sector is of canonical Einstein-Hilbert form, and we may instead have direct couplings of the field $X$ to the matter sector.  While calculations in the Jordan frame are complicated by the need to treat gravity dynamically, those in the Einstein frame are complicated by the Weyl rescaling itself and the need to rescale the matter fields, which must be done on a model-by-model basis, as we will now describe.

	The Weyl rescaling takes the form
	\begin{equation}
	    \label{eq:Weyl_rescaling}
		{g}_{\mu \nu}=A^2(X)\tilde{g}_{\mu \nu},
	\end{equation}
	where $\tilde{g}_{\mu\nu}$ is the Einstein-frame metric, $A^2(X)=\tilde M_{\rm Pl}^2F^{-1}(X)$ is the squared coupling function and $\tilde M_{\rm Pl}$ is the reduced Planck mass of the Einstein frame.
	After some algebra, the transformation~\eqref{eq:Weyl_rescaling} leads to the Einstein-frame action
	\begin{equation}
	    \label{eq:EF_action_general}
		S=\int \D^4{x} \sqrt{-\tilde{g}} \left[\frac{1}{2}\tilde M_{\rm Pl}^2\tilde{R} - \frac{1}{2}\tilde{g}^{\mu\nu}\partial_\mu \tilde{\chi} \partial_\nu \tilde{\chi} - \tilde{A}^4(\tilde{\chi})\tilde{U}(\tilde{\chi})\right] +  S_{\rm{m}}[ \tilde{A}^2(\tilde{\chi})\tilde{g}_{\mu \nu},\{\psi\}],	
	\end{equation}
	where $\tilde R$ is the Ricci scalar built with the Einstein-frame metric $\tilde{g}_{\mu\nu}$. In addition, $\tilde{A}(\tilde{\chi})\equiv A(X(\tilde{\chi}))$ and $\tilde{U}(\tilde{\chi})\equiv U(X(\tilde{\chi}))$ are respectively the coupling function and potential expressed in terms of the canonically normalized field 
	\begin{equation}
		\tilde{\chi}(X)\equiv \tilde M_{\rm Pl}\int^{X}_{X_0}\D{\hat{X}} \sqrt{\frac{{Z}(\hat{X})}{{F}(\hat{X})}+3[{F'}(\hat{X})]^2},
		\label{O2}
	\end{equation}
	where the prime in $F'(\hat{X})$ indicates the derivative with respect to the argument. We see from Eq.~\eqref{eq:EF_action_general} that, in the Einstein frame, matter fields still move on geodesics of the Jordan-frame metric, potentially giving rise to deviations from Einstein gravity that can be interpreted as a fifth force due to the exchange of fluctuations in the $\tilde{\chi}$ field.
 
 If $\tilde{A}^2$ can be expanded about unity, such that $\tilde{A}^2(\tilde{\chi})=1+\frac{1}{n!}\frac{\tilde{\chi}^n}{M^n}+\dots$, we can similarly expand the matter action to show that
\begin{equation}
    \label{eq:actionexpansion}
    S_{\rm{m}}[\tilde{A}^{2}(\tilde{\chi}) \tilde{g}_{\mu \nu},\{\psi\}]= S_{\rm{m}}\left[\tilde{g}_{\mu \nu},\{\psi\}\right]+\frac{1}{n!}T\frac{\tilde{\chi}^{n}}{M^{n}}+\dots,
\end{equation}
wherein we see that the field $\tilde{\chi}$ couples to the trace of the energy-momentum tensor of the matter sector $T=\tilde{g}_{\alpha\beta}T^{\alpha\beta}$. Expressing the fifth-force coupling in this way, however, obscures the role that scale-symmetry breaking plays in the emergence of fifth forces~\cite{Burrage:2018dvt}, as we will describe in the next subsection.


\subsection{Fifth forces in the Einstein frame}
	
	The fifth forces that arise in scalar-tensor theories can be studied in multiple ways. For instance, one can solve the classical equations of motion and extract the corrections to the Newtonian potential from the spatial gradient of the solution (see, e.g., Ref.~\cite{Joyce:2014kja, Burrage:2017qrf}). Equivalently, we can work with the tree-level matrix element for the $t$-channel exchange of the scalar mediator, from which we can then extract the non-relativistic potential (see, e.g., Refs.~\cite{Burrage:2018dvt, Banks:2020gpu}).
	
	We consider the following toy model, as constructed in Ref.~\cite{Burrage:2018dvt} and written in terms of the Jordan-frame metric $g_{\mu \nu}$:
	\begin{align}
			S_{\rm{m}}=\int \D^4{x} \sqrt{-g}& \left[-\frac{1}{2}g^{\mu \nu}\partial_\mu \Phi \partial_\nu \Phi -\frac{1}{2}g^{\mu \nu}\partial_\mu \Theta \partial_\nu \Theta -\bar{\psi}i{\stackrel{\leftrightarrow}{\slashed{\partial}}}\psi - y\bar{\psi}\Phi\psi\right.\nonumber\\ &\qquad \left.-U(\Phi, \Theta)+\frac{1}{2} \mu_{\theta}^{2} A^{-2}(X) \Theta^{2}-\frac{\lambda_{\theta}}{4 !} \Theta^{4}-\frac{3}{2} \frac{\mu_{\theta}^{4}}{\lambda_{\theta}} A^{-4}(X)\right],
		\label{sm}
	\end{align}
	where
		\begin{equation}
			U(\Phi, \Theta)=\frac{\lambda}{4 !}\left(\Phi^{2}-\frac{\beta}{\lambda} \Theta^{2}\right)^{2}-\frac{1}{2} \mu^{2}\left(\Phi^{2}-\frac{\beta}{\lambda} \Theta^{2}\right)+\frac{3}{2} \frac{\mu^{4}}{\lambda} .\label{comb_pot}
	\end{equation}
  The matter sector contains a would-be Higgs field $\Phi$ and a Dirac fermion $\psi$ (a proxy for the SM electron). The fermion kinetic term has been antisymmetrized, with ${\stackrel{\leftrightarrow}{\slashed{\partial}}}\equiv e^{\mu}_{a}\gamma^a{\stackrel{\rightarrow}{\partial}}_\mu-{\stackrel{\leftarrow}{\partial_\mu}} e^{\mu}_{a}\gamma^a$ (where we make use of the vierbein $e^{\mu}_{a}$), so that we can omit the spin connection from the action (see refs.~\cite{Burrage:2018dvt, Ferreira:2016kxi}). These two fields interact through a Yukawa coupling, which gives mass to the fermion after the would-be Higgs sector undergoes symmetry breaking. The additional $\Theta$ field has been introduced so that we can move smoothly between two scenarios of scale breaking, and independently of the non-minimally coupled field $X$ and its dynamics:~the first ($\beta \to 0$) in which the scale breaking is explicit, due to the presence of the dimensionful parameter $\mu$ in the potential of the matter fields; the second ($\mu \to 0$) in which the scale breaking is dynamical, arising along a particular trajectory in the $\Phi-\Theta$ field space without explicit scale-breaking terms appearing in the Lagrangian. The specific choice of couplings between the $\Theta$ and $X$ fields has been tuned so that these fields do not have a mass mixing in the Einstein frame [see, e.g., Eq.~\eqref{eq:EFrameL_with_theta} below], while also allowing us to establish a hierarchy between the masses of the three physical modes (see Ref.~\cite{Burrage:2018dvt}).
	  
	  The two limiting cases for the distinct sources of scale breaking are as follows:
\paragraph*{\bf Pure explicit breaking (prototype SM Higgs sector) $\boldsymbol{\beta\to 0}$:}
	    In the limit in which $\beta\to0$, the mixings between $\Phi$ and $\Theta$ vanish, decoupling $\Theta$ from the matter Lagrangian. We are then left with the following potential in the Jordan frame [Eq.~\eqref{comb_pot}]:
	    \begin{equation}
	        U(\Phi)=\frac{\lambda}{4 !}\Phi^{4}-\frac{1}{2} \mu^{2}\Phi^{2}+\frac{3}{2} \frac{\mu^{4}}{\lambda},
	    \end{equation}
	    which is just the prototype of the SM Higgs potential, where we can see that the constant term in the potential $U$ ensures that the vacuum has zero energy density in the symmetry-broken phase of the would-be Higgs sector. In this case, the term quadratic in $\Phi$ provides an explicit source of scale breaking through the dimensionful mass parameter $\mu$. As we will show below, it is this term that plays the key role in allowing fifth forces to couple to the fermion field $\psi$.
	    
\paragraph*{\bf Pure dynamical scale breaking (prototype Higgs-dilaton model) $\boldsymbol{\mu\to0}$:}
	    In this limit, all the sources of explicit scale breaking vanish from $U(\Phi,\Theta)$, leaving a scale-invariant potential. We therefore do not expect the conformal field $X$ to couple to this potential in the Einstein frame, leaving the fermionic sector free of fifth forces. The potential is reduced to 
	    \begin{equation}
		U(\Phi, \Theta)=\frac{\lambda}{4 !}\left(\Phi^{2}-\frac{\beta}{\lambda} \Theta^{2}\right)^{2},
	    \end{equation}
	    and analogous potentials appear in Higgs-dilaton theories~\cite{Wetterich:1987fm, Buchmuller:1988cj, Shaposhnikov:2008xb, Shaposhnikov:2008xi, Blas:2011ac, Garcia-Bellido:2011kqb, Garcia-Bellido:2012npk, Bezrukov:2012hx, Henz:2013oxa, Rubio:2014wta, Karananas:2016grc, Ferreira:2016vsc, Ferreira:2016kxi, Casas:2017wjh, Ferreira:2018qss}. In those scenarios, however, both $\Phi$ and $\Theta$ are non-minimally coupled to the Ricci scalar in the Jordan frame (and the dilaton is the light degree of freedom with the potential to mediate long-range forces), whereas we will take only the additional field $X$ to be non-minimally coupled.
	   The classical scale symmetry of this model is broken when the scalar fields $\Phi$ and $\Theta$ obtain non-vanishing vevs, leading to the emergence of a scale. 
	   Since this scale appears indirectly through the stabilization of the fields, we will refer to this as dynamical scale breaking~\cite{Ferreira:2018itt}.
	   In contrast to the case of explicit scale breaking, the fifth force mediated by $X$ will not couple to the matter fields $\psi$.  In what follows, we will show this by explicit calculation of the matrix elements. Alternatively, this can be understood in terms of the existence of a conserved dilatation current (see, e.g., Ref.~\cite{Ferreira:2016kxi, Brax:2014baa}).

    Let us now turn our attention to an explicit calculation of the fifth forces for this model. First, we need to express the matter action in terms of the Einstein-frame metric $\tilde{g}_{\mu \nu}$. To do so, we must perform the conformal transformation defined previously in Eq.~(\ref{eq:Weyl_rescaling}). This  gives 
		\begin{align}
			S_{\rm{m}}&=\int \D^4{x} \sqrt{-\tilde{g}} \left[-\frac{1}{2}\tilde{A}^2(\tilde{\chi})\tilde{g}^{\mu \nu}\partial_\mu \Phi \partial_\nu \Phi -\frac{1}{2}\tilde{A}^2(\tilde{\chi})\tilde{g}^{\mu \nu}\partial_\mu \Theta \partial_\nu \Theta-\tilde{A}^3(\tilde{\chi})\bar{\psi}i\stackrel{\leftrightarrow}{\tilde{\slashed{\partial}}}\psi \right.\nonumber\\ &\qquad \left.- y \tilde{A}^4(\tilde{\chi})\bar{\psi}\Phi\psi-\tilde{A}^4(\tilde{\chi})U(\Phi, \Theta)+\frac{1}{2}\tilde{A}^2(\tilde{\chi}) \mu_{\theta}^{2} \Theta^{2}-\frac{\lambda_{\theta}}{4 !} \tilde{A}^4(\tilde{\chi})\Theta^{4}-\frac{3}{2} \frac{\mu_{\theta}^{4}}{\lambda_{\theta}} \right],
		\label{sm1}
	\end{align}
    where $\tilde{\chi}\equiv \tilde{\chi}(X)$ is the canonically normalized field [cf.~Eq.~\eqref{O2}] and $\stackrel{\leftrightarrow}{\tilde{\slashed{\partial}}}\equiv \tilde{e}^{\mu}_{a}\gamma^a{\stackrel{\rightarrow}{\partial}}_\mu-{\stackrel{\leftarrow}{\partial_\mu}} \tilde{e}^{\mu}_{a}\gamma^a=\tilde{A}(\tilde{\chi})e^{\mu}_{a}\gamma^a{\stackrel{\rightarrow}{\partial}}_\mu-{\stackrel{\leftarrow}{\partial_\mu}} e^{\mu}_{a}\gamma^a\tilde{A}(\tilde{\chi})$.	To leave the matter sector as close to being canonically normalized as possible, we redefine the fields according to their classical scaling dimensions, such that
	\begin{align}
		\tilde{\phi} \equiv \tilde{A}(\tilde{\chi}) \Phi, \qquad\tilde{\theta} \equiv \tilde{A}(\tilde{\chi}) \Theta, \qquad \tilde{\psi} \equiv \tilde{A}^{3 / 2}(\tilde{\chi}) \psi.
		\label{scasca}
	\end{align}
	With this, the Lagrangian becomes
	\begin{align}
			\tilde{\mathcal{L}}_{\rm{m}}=& -\frac{1}{2} \tilde{g}^{\mu \nu} \partial_{\mu} \tilde{\phi} \partial_{\nu} \tilde{\phi}+\tilde{g}^{\mu \nu} \tilde{\phi} \partial_{\mu} \tilde{\phi} \partial_{\nu} \ln \tilde{A}(\tilde{\chi})-\frac{1}{2} \tilde{g}^{\mu \nu} \tilde{\phi}^{2} \partial_{\mu} \ln \tilde{A}(\tilde{\chi}) \partial_{\nu} \ln \tilde{A}(\tilde{\chi}) \nonumber\\
			& -\frac{1}{2} \tilde{g}^{\mu \nu} \partial_{\mu} \tilde{\theta} \partial_{\nu} \tilde{\theta}+\tilde{g}^{\mu \nu} \tilde{\theta} \partial_{\mu} \tilde{\theta} \partial_{\nu} \ln \tilde{A}(\tilde{\chi})-\frac{1}{2} \tilde{g}^{\mu \nu} \tilde{\theta}^{2} \partial_{\mu} \ln \tilde{A}(\tilde{\chi}) \partial_{\nu} \ln \tilde{A}(\tilde{\chi}) \nonumber\\
			& +\tilde{U}(\tilde{\phi},\tilde{\theta},\tilde{\chi})-\bar{\tilde{\psi}}i\stackrel{\leftrightarrow}{\tilde{\slashed{\partial}}}\tilde{\psi}-y \bar{\tilde{\psi}} \tilde{\phi} \tilde{\psi}-\frac{1}{2} \mu_{\theta}^{2} \tilde{\theta}^{2}+\frac{\lambda_{\theta}}{4 !} \tilde{\theta}^{4}+\frac{3}{2} \frac{\mu_{\theta}^{4}}{\lambda_{\theta}},\label{mat_lag}
	\end{align}
	where
		\begin{align}
		    \label{eq:EFrameL_with_theta}
			\tilde{U}(\tilde{\phi}, \tilde{\theta}, \tilde{\chi})=\frac{\lambda}{4 !}\left(\tilde{\phi}^{2}-\frac{\beta}{\lambda} \tilde{\theta}^{2}\right)^{2}-\frac{1}{2} \mu^{2}\tilde{A}^2(\tilde{\chi})\left(\tilde{\phi}^{2}-\frac{\beta}{\lambda} \tilde{\theta}^{2}\right)
			+\frac{3}{2} \tilde{A}^4(\tilde{\chi})\frac{\mu^{4}}{\lambda}.
	\end{align}
	Thus, we can see that the redefinitions from Eq.~(\ref{scasca}) eliminate all the couplings of $\tilde{\chi}$ in the fermionic sector and in the pure $\Theta$ potential from the last line of Eq.~\eqref{mat_lag}. However, the same does not apply to $\tilde{U}(\tilde{\phi}, \tilde{\theta}, \tilde{\chi})$, since it contains dimensionful parameters. Moreover, as explained before, the only terms coupling to $\tilde{\chi}$ in the Einstein frame are the ones which break the scale symmetry explicitly. 
	
	Keeping the calculation as generic as possible, we now expand the coupling function as
	\begin{align}
		\tilde{A}^{2}(\tilde{\chi})=a+b \frac{\tilde{\chi}}{\tilde{M}}+c \frac{\tilde{\chi}^{2}}{\tilde{M}^{2}}+\mathcal{O}\left(\frac{\tilde{\chi}^{3}}{\tilde{M}^{3}}\right),
		\label{para}
	\end{align}
	where $\tilde{M}$ is an energy scale, and $a$, $b$ and $c$ are dimensionless constants, which will be defined for specific models. After including the original kinetic energy term for $\tilde{\chi}$ from Eq.~\eqref{eq:EF_action_general} and making a further redefinition $\partial_\mu \tilde{\chi} \to \sqrt{1+\tilde{\theta}^2 + \tilde{\phi}^2}\; \partial_{\mu} \ln \tilde{A}(\tilde\chi)$, the non-gravitational part of the Einstein-frame Lagrangian  can be written up to second order in $\tilde{M}^{-1}$ as
	\begin{align}
			\tilde{\mathcal{L}}=& -\frac{1}{2} \tilde{g}^{\mu \nu} \partial_{\mu} \tilde{\chi} \partial_{\nu} \tilde{\chi}-\frac{1}{2} \tilde{g}^{\mu \nu} \partial_{\mu} \tilde{\phi} \partial_{\nu} \tilde{\phi}+\frac{1}{2} \tilde{g}^{\mu \nu} \frac{\tilde{\phi}}{\tilde{M}}\left(b+2 a c \frac{\tilde{\chi}}{\tilde{M}}-b^{2} \frac{\tilde{\chi}}{2 \tilde{M}}\right) \partial_{\mu} \tilde{\phi} \partial_{\nu} \tilde{\chi} \nonumber\\
			-&\frac{1}{2} \tilde{g}^{\mu \nu} \partial_{\mu} \tilde{\theta} \partial_{\nu} \tilde{\theta}+\frac{1}{2} \tilde{g}^{\mu \nu} \frac{\tilde{\theta}}{\tilde{M}}\left(b+2 a c \frac{\tilde{\theta}}{\tilde{M}}-b^{2} \frac{\tilde{\chi}}{2 \tilde{M}}\right) \partial_{\mu} \tilde{\theta} \partial_{\nu} \tilde{\chi} -\tilde{U}(\tilde{\phi}, \tilde{\theta}, \tilde{\chi})\nonumber\\
			-&\bar{\tilde{\psi}} i\stackrel{\leftrightarrow}{\tilde{\slashed{\partial}}}\tilde{\psi}-y \bar{\tilde{\psi}} \tilde{\phi} \tilde{\psi}-\frac{1}{2} \mu_{\theta}^{2} \tilde{\theta}^{2}+\frac{\lambda_{\theta}}{4 !} \tilde{\theta}^{4}+\frac{3}{2} \frac{\mu_{\theta}^{4}}{\lambda_{\theta}}\cdots,
		\label{genein}
	\end{align}
	where 
	\begin{align}
			\tilde{U}(\tilde{\phi}, \tilde{\theta}, \tilde{\chi})=&\frac{\lambda}{4 !}\left(\tilde{\phi}^{2}-\frac{\beta}{\lambda} \tilde{\theta}^{2}\right)^{2}-\frac{1}{2} \mu^{2}\left(\tilde{\phi}^{2}-\frac{\beta}{\lambda} \tilde{\theta}^{2}\right)\left(a+b \frac{\tilde{\chi}}{\tilde{M}}+c \frac{\tilde{\chi}^{2}}{\tilde{M}^{2}}\right) \nonumber\\
			+&\frac{3}{2} \frac{\mu^{4}}{\lambda}\left(a+2ab \frac{\tilde{\chi}}{\tilde{M}}+(2ac+b^2) \frac{\tilde{\chi}^{2}}{\tilde{M}^{2}}\right).
			\label{potential_comb}
	\end{align}
	
	We see that there are both kinetic and mass mixings of $\tilde{\phi}$ and $\tilde{\theta}$ with $\tilde{\chi}$. However, fifth forces arising through kinetic mixings when these involve a field with non-zero mass are suppressed due to the additional momentum dependence ($\propto q^2$) that occurs for each insertion into the matrix element of the kinetic mixing operator. 
	As a result, the mass mixing will provide the dominant fifth force. 
	Moreover, since we are free to choose $\mu_{\theta}^{2} \gg \mu^{2}$, the $\tilde{\theta}$ field can be decoupled from the long-range fifth forces. We can then focus solely on the mixing between the would-be Higgs field $\tilde{\phi}$ and the conformally coupled scalar $\tilde{\chi}$, as we will do in the next subsection.

\subsection{Fifth forces}
	
	As a concrete example, we will now specialize to the Brans-Dicke theory \cite{Brans:1961sx}, whose Jordan-frame action is
	\begin{equation}\label{BDactionX}
		S=\int \mathrm{d}^{4} x \sqrt{-g}\left[\frac{X}{2} R-\frac{\omega(X)}{2 X} g^{\mu \nu} \partial_{\mu} X \partial_{\nu} X\right]+S_{\rm{m}}[g_{\mu \nu}, \{\psi\}] .
	\end{equation}
	After performing the conformal transformation to the Einstein frame and canonically normalizing the fields (the explicit calculation can be found in Ref.~\cite{Burrage:2018dvt}), we find that the coupling function takes the form
	\begin{align}
		A^{2}(X(\tilde{\chi}))=&\frac{M_{\rm{Pl}}^2}{X}=\exp\left[2\frac{\tilde{\chi}}{\tilde{M}}\right],
		\label{parcham}
	\end{align}
wherein we have taken $\omega(X)=\omega$ to be a constant and defined
	\begin{equation}
	    \tilde{M}^2=2(2\omega +3)\tilde{M}_{\rm{Pl}}^2.
	\end{equation}
	Thus, the Brans-Dicke model amounts to taking $a=1$, $b=2$ and $c=2$ in Eq.~(\ref{para}). It then follows from Eq.~(\ref{potential_comb}) that the Einstein-frame potential is
		\begin{align}
			\tilde{U}(\tilde{\phi}, \tilde{\theta}, \tilde{\chi})=&\frac{\lambda}{4 !}\left(\tilde{\phi}^{2}-\frac{\beta}{\lambda} \tilde{\theta}^{2}\right)^{2}-\frac{1}{2} \mu^{2}\left(\tilde{\phi}^{2}-\frac{\beta}{\lambda} \tilde{\theta}^{2}\right)\left(1+2 \frac{\tilde{\chi}}{\tilde{M}}+2 \frac{\tilde{\chi}^{2}}{\tilde{M}^{2}}\right) \nonumber\\
			+&\frac{3}{2} \frac{\mu^{4}}{\lambda}\left(1+4 \frac{\tilde{\chi}}{\tilde{M}}+8 \frac{\tilde{\chi}^{2}}{\tilde{M}^{2}}\right)-\frac{1}{2} \mu_{\theta}^{2} \tilde{\theta}^{2}+\frac{\lambda_{\theta}}{4 !} \tilde{\theta}^{4}+\frac{3}{2} \frac{\mu_{\theta}^{4}}{\lambda_{\theta}}.
	\end{align}
	
	The fields acquire the vevs
	\begin{align}
		&v_{\tilde{\phi}}=\pm\left(\frac{6 \mu^{2}+\beta v_{\tilde{\theta}}^{2}}{\lambda}\right)^{1 / 2},&  &v_{\tilde{\theta}}=\pm'\left(\frac{6 \mu_{\theta}^{2}}{\lambda_{\theta}}\right)^{1 / 2},&  &v_{\chi}=0,&
		\label{vevss}
	\end{align}
	where the $\prime$ indicates that the choice of sign for the two non-vanishing vevs is independent. Expanding around the vevs ($\tilde{\phi} \rightarrow v_{\tilde{\phi}}+\tilde{\phi}$, $\tilde{\theta} \rightarrow v_{\tilde{\theta}}+\tilde{\theta}$ and $\tilde\chi \rightarrow v_{\tilde{\chi}}+\tilde{\chi}$) and recalling that the main contribution to the fifth forces is given by the mass mixing between $\tilde{\phi}$ and $\tilde{\chi}$, the operator of interest from Eq.~\eqref{genein} is given by
	\begin{equation}
	    \label{eq:massmixing}
		\tilde{\mathcal{L}} \supset \alpha_{\rm{M}} \tilde\phi \tilde\chi=2 \mu^{2} \frac{v_{\tilde{\phi}}}{\tilde{M}} \tilde{\phi} \tilde{\chi}.
	\end{equation}
We are now in the position to calculate the matrix elements for the fifth force.

\begin{figure}[t]
	    \centering
	\includegraphics[scale=0.28]{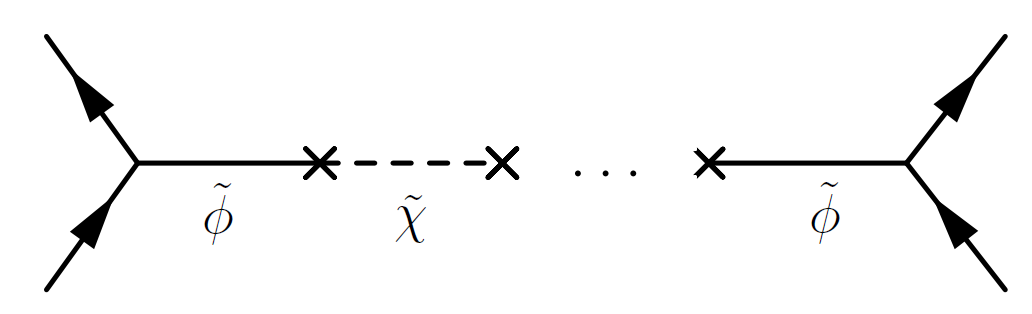}
	    \caption{Diagrammatic representation of the infinite series of diagrams contributing to the M\o{}ller scattering in the Einstein frame.}
	    \label{fig:massmixing}
	\end{figure}

	We proceed by considering the scalar contributions to the M\o{}ller scattering ($e^{-}e^{-} \to e^{-}e^{-}$) for our fermion $\psi$. These arise from the series of diagrams shown in Fig.~\ref{fig:massmixing}. The external fermions couple only to the would-be Higgs field, represented by a continuous line, which then oscillates into a $\tilde \chi$ particle (dashed line) via the mass mixing term from the effective Lagrangian [Eq.~\eqref{eq:massmixing}]. The ellipsis represents the infinite series of insertions of the mass mixing.

	The resulting matrix element is given by
		\begin{align}
			i \mathcal{M}\left(e^{-} e^{-} \rightarrow e^{-} e^{-}\right) &\supset \bar{u} \left(\mathbf{p}_{1}, s_{1}\right)(-i y) u\left(\mathbf{p}_{3}, s_{3}\right)\nonumber \\
			& \times \frac{i}{t-m_{\tilde{\phi}}^{2}}\left[\sum_{n=0}^{\infty}\left(i \alpha_{\rm{M}}\right)^{2 n}\left(\frac{i}{t-m_{\tilde{\phi}}^{2}}\right)^{n}\left(\frac{i}{t}\right)^{n}\right] \nonumber\\
			& \times \bar{u}\left(\mathbf{p}_{2}, s_{2}\right)(-i y) u\left(\mathbf{p}_{4}, s_{4}\right).
        \end{align}
	Since we assume the scattering fermions to be distinguishable, we need only consider the $t$-channel exchange, where $t = -(p_1 - p_3)^2$ is the usual Mandelstam variable. Also, $u(\mathbf{p},s)$ and $\bar{u}(\mathbf{p},s)$ are respectively the Dirac four-spinor and its Dirac conjugate, with spin projection $s$. The resulting non-relativistic potential is given by
		\begin{equation}
			\tilde{V}(r) =-y^{2} \int \frac{\mathrm{d}^{3} \mathbf{Q}}{(2 \pi)^{3}} e^{i \mathbf{Q} \cdot \mathbf{x}} \frac{\mathbf{Q}^{2}}{\mathbf{Q}^{2}\left(\mathbf{Q}^{2}+m_{\tilde{\phi}}^{2}\right)-\alpha_{\rm{M}}^{2}}
			\approx-\frac{y^{2}}{4 \pi}\left(1-\frac{\alpha_{\rm{M}}^{2}}{m_{\tilde{\phi}}^{4}}\right) \frac{e^{-m_{\rm{h}} r}}{r}-\frac{y^{2}}{4 \pi} \frac{\alpha_{\rm{M}}^{2}}{m_{\tilde{\phi}}^{4}} \frac{1}{r},
		\end{equation}
	where $m_{\rm{h}}$ is the mass of the would-be Higgs boson and the potential has been expanded to leading order in $\alpha^2_{\rm{M}}$. 
	Isolating the fifth-force contribution and plugging in $\alpha_{\rm{M}}$, as extracted from Eq.~\eqref{eq:massmixing}, we find
	\begin{equation}
		\tilde{V}_5(r) =-\frac{1}{4 \pi r} \frac{m_{e}^{2}}{\tilde M^2_{\rm{Pl}}2(2\omega+3)} \frac{4\mu^4}{m^4_{\tilde{\phi}}},	\label{solu}
	\end{equation}
	where we have chosen the fermions to represent electrons with mass $m_e$. Notice that, since the fifth-force mediator is massless, the potential has a similar form to the usual Newtonian gravitational potential.
	
	To study how the different mechanisms of scale breaking affect the modification to the Yukawa potential [Eq.~\eqref{solu}], we need only recall that the mass of the $\tilde{\phi}$ field is given by
    \begin{equation}
		m^2_{\tilde{\phi}}=2\mu^2 + \frac{\beta v_{\tilde{\theta}}^2}{3}.
	\end{equation}
 
    \paragraph*{\bf Pure explicit scale breaking (SM toy model) $\boldsymbol{\beta\to 0}$:} In this limit, the mass of the $\tilde{\phi}$ field reduces to
	    \begin{equation}
	        m^2_{\tilde{\phi}}=2\mu^2,
	    \end{equation} 
	    agreeing with the numerator of the fraction in the potential~\eqref{solu}. Hence, the modification to the Yukawa potential becomes independent of the Higgs mass, and we find
	    \begin{equation}
		\tilde{V}_5(r) =-\frac{1}{4 \pi r} \frac{m_{e}^{2}}{\tilde M^2_{\rm{Pl}}2(2\omega+3)}.
		\label{exp_eins}
	\end{equation}Such a contribution to the non-relativistic potential can lead to significant deviations in the inferred gravitational force.  A comprehensive review on the different tests and bounds on fifth forces can be found in Ref.~\cite{Burrage:2017qrf}, where the authors mainly focus on the chameleon model, which applies to our results given that we consider only up to second order in the field fluctuations. In particular, the most stringent constraint  at Solar System scales is given by the Cassini spacecraft~\cite{Bertotti:2003rm}, setting a bound on $\omega \gg40,000$. Bounds at cosmological scales are less stringent\footnote{Even though Solar System scale tests are more constraining than cosmological ones, they are more affected by higher-order terms, making it possible to avoid the bounds through screening mechanisms.}, such as those based on  Cosmic Microwave Background data from Planck~\cite{Avilez:2013dxa}, which are consistent with $\omega>692$ at the $99\%$ confidence level. Therefore, in the absence of any screening mechanism, we can see that a fine tuning of the value of $\omega$ in the case of  pure explicit scale breaking is necessary to achieve an agreement with experiments.
	
\paragraph*{\bf Pure dynamical scale breaking (Higgs-dilaton model) $\boldsymbol{\mu\to0}$:}
	    In this case, the numerator of the modified Yukawa potential [Eq.~\eqref{solu}] tends to zero, whereas the denominator tends to
	    \begin{equation}
	        m^2_{\tilde{\phi}}=\frac{\beta v_{\tilde{\theta}}^2}{3}.
	    \end{equation}
	   Hence, even though classically scale-invariant theories might break the scale symmetry dynamically, the fifth forces still do not couple to the fermionic sector. It is important to remark that  the vev of the would-be Higgs field $\tilde{\phi}$ field [Eq.~(\ref{vevss})] does not vanish in the limit $\mu\to0$, such that the mass-generation mechanism for the elementary fermions is preserved (with $m_e=yv_{\tilde{\phi}}$). More generally, we see that the fifth-force coupling is proportional to the ratio $\mu/m_{\tilde{\phi}}$, such that we can suppress fifth forces by combining explicit and dynamical scale-breaking mechanisms~\cite{Burrage:2018dvt}.
	
For this tree-level example, the transformation to the Einstein frame and the subsequent calculation of the matrix elements were easily tractable. This may not be the case, in general, however. In the next section, we will describe in detail how we can proceed directly in the Jordan frame (or Jordan-like frames), without performing the conformal transformation and subsequent field rescalings.


\section{Linearized scalar-tensor gravity}\label{thirdS}
	
	If we wish to proceed directly in the Jordan frame, it is necessary to treat the metric dynamically and to work with linearized gravity, as we can in the weak-field limit. We will first review the linearization of Einstein gravity, which has been widely studied (see, e.g., Refs.~\cite{Fierz:1939ix,Donoghue:1995cz,Donoghue:2017pgk}). We will then generalize this procedure to actions with non-minimal couplings of Brans-Dicke type, taking care to consider the differing symmetries of this class of scalar-tensor theories.

	
\subsection{Standard gravity}
	
	Gauge symmetries reflect the redundancies of a theory, so we need to break these symmetries through a gauge fixing term to fully describe the field theory we are working with. We therefore take the Einstein-Hilbert action to be of the form
	\begin{equation}
		S_{\rm EH}=\int \D^4{x} \sqrt{-g} \left[ \frac{{M_{\rm{Pl}}}^2}{2}R + \lgr_{\rm{gf}}\right],
		\label{two}
	\end{equation}
	where we have included the gauge fixing terms via $\lgr_{\rm{gf}}$.
	
	As we will justify later, we choose to work with the harmonic gauge, since it can be expressed at the level of the Lagrangian. The harmonic gauge is defined such that
	\begin{equation}
	    \label{HG-constraint}
	    \nabla^\mu\nabla_\mu=\partial^\mu\partial_\mu,
	\end{equation}
	where $\nabla_\mu$ is the usual covariant derivative in General Relativity. This constraint is uniquely satisfied by setting
	\begin{equation}\label{HG_condition}
	    \Gamma^\mu=g^{\alpha \beta}\Gamma^\mu_{\alpha \beta}=0,
	\end{equation}
	where $\Gamma^\mu_{\alpha \beta}$ are the Christoffel symbols. Therefore, the gauge fixing term for the harmonic gauge becomes
	\begin{equation}
		\lgr_{\rm{gf}}=-\frac{{M_{\rm{Pl}}}^2}{4}g_{\alpha \beta}\Gamma^\alpha\Gamma^\beta.
		\label{gf}
	\end{equation}
	
	To linearize gravity, we must make small perturbations of the metric around a constant background. In this paper, we will assume this background to be flat, such that
	\begin{equation}
		g_{\mu \nu}=\eta_{\mu \nu} + h_{\mu \nu},
	\end{equation}
	where $\eta_{\mu\nu}={\rm diag}(-1,1,1,1)$ is the Minkowski metric. We then arrive at the following weak-field expansions:
	\begin{subequations}
	\begin{align}
		\sqrt{-g}\approx&\, 1 + \frac{1}{2}\eta_{\mu \nu}h^{\mu \nu}, \\
		R_{\mu \nu}^{(1)}=&\,\frac{1}{2}\left( \partial^\rho \partial_\mu h_{\nu \rho} + \partial^\rho \partial_\nu h_{\mu \rho} -\Box h_{\mu \nu} -\partial_\mu \partial_\nu h\right), \label{E1}\\
			R_{\mu \nu}^{(2)}=&\,\frac{1}{2}h^{\rho \sigma} \partial_\mu \partial_\nu h_{\rho \sigma} -h^{\rho \sigma} \partial_\mu {\partial_{(\nu}} h_{\rho) \sigma}	+\frac{1}{4}\partial_\mu h^{\rho \sigma} \partial_\nu h_{\rho \sigma}\nonumber \\
			&\,+\partial^\sigma h^{\rho}_{\nu} {\partial_{[\sigma}} h_{\rho] \mu} + \frac{1}{2}\partial_\sigma(h^{\sigma \rho}\partial_\rho h_{\mu \nu}) - \frac{1}{4}\partial^\rho h \partial_{\rho} h_{\mu \nu} \nonumber\\
			&\,-(\partial_\sigma h^{\sigma \rho} - \frac{1}{2}\partial^\rho h){\partial_{(\mu}} h_{\nu) \rho}, \\
		\Gamma^{\mu(1)}_{\alpha \beta} =&\, \frac{1}{2}\eta^{\mu \lambda}(\partial_\alpha h_{\lambda \beta} + \partial_\beta h_{\alpha \lambda} - \partial_\lambda h_{\alpha \beta}),\label{E2}
	\end{align}
	\end{subequations}
	where the exponent in parenthesis shows the order in the metric fluctuations $h$. With these ingredients, we can then determine the expansion of the terms in Eq.~(\ref{two}) up to second order:
	\begin{subequations}
	\begin{align}
		\sqrt{-g}\frac{{M_{\rm{Pl}}}^2}{2}R=&\frac{{M_{\rm{Pl}}}^2}{2}\left(1 + \frac{1}{2}\eta_{\mu\nu}h^{\mu\nu}\right)\left[R_{\mu\nu}(\eta^{\mu\nu} -h^{\mu\nu})\right]\nonumber\\
		=&\frac{{M_{\rm{Pl}}}^2}{2}\left(R^{(1)} - R_{\mu\nu}^{(1)}h^{\mu \nu} + \frac{1}{2}R^{(1)}\eta_{\mu \nu}h^{\mu \nu}+R^{(2)}\right), \\
		\sqrt{-g}\lgr_{\rm{gf}}=&-\sqrt{-g}\frac{{M_{\rm{Pl}}}^2}{4}g_{\alpha \beta}\Gamma^\alpha\Gamma^\beta = -\frac{{M_{\rm{Pl}}}^2}{4}\eta_{\alpha \beta}\eta^{\mu\nu}\Gamma^{\alpha(1)}_{\mu \nu}\eta^{\sigma \rho}\Gamma^{\beta(1)}_{\sigma \rho}, \\
		\sqrt{-g}\lgr_{\rm{m}}(g_{\mu \nu})=&\frac{1}{2}h^{\mu \nu}T_{\mu \nu} + \lgr_{\rm{m}}(\eta_{\mu \nu})\label{R2},
	\end{align}
	\end{subequations}
	where we have re-introduced the matter sector, giving rise to the contributions from its energy-momentum tensor $T_{\mu\nu}$.
	
	 Making use of the results in Eqs.~(\ref{E1}) to (\ref{E2}) and integrating by parts, we obtain the following expression for the Lagrangian up to second order in $h_{\mu \nu}$:
	\begin{equation}
		\lgr=\frac{M_{\rm{Pl}}^2}{4}\left[ \frac{1}{4}\partial_\mu h \partial^\mu h -\frac{1}{2}\partial_\rho h_{\mu \nu}\partial^\rho h^{\mu \nu}\right] + \frac{M_{\rm{Pl}}^2}{2}\left[\partial^\mu\partial^\nu h_{\mu \nu} - \Box h\right] +\frac{1}{2}h^{\mu\nu}T_{\mu\nu} + \lgr_{\rm{m}}(\eta_{\mu\nu}).
	\end{equation}
	The first term corresponds to the kinetic energy of the graviton and can be canonically normalized by rescaling it such that $h_{\mu \nu} \to 2h_{\mu \nu}/M_{\rm{Pl}}$. We draw attention to the second term, containing second derivatives of the graviton field, which can be removed on integrating by parts and setting the boundary terms to zero. In contrast, the analogous term for the scalar-tensor theory will not vanish, and it will be the source of the fifth forces in the Jordan frame, as we will show in the next subsection.

	
\subsection{Jordan frame}
	
	We now repeat the linearization for Brans-Dicke-type scalar-tensor theories. The action in this case will have a non-minimal coupling to the Ricci scalar, such that
	\begin{equation}\label{Action-BD-GF}
		S=\int \D^4{x} \sqrt{-g} \left[\frac{F(X)}{2}R + \lgr'_{\rm{gf}} - \frac{Z(X)}{2}\partial_\mu X \partial^\mu X - U(X)\right] +  S_{\rm{m}}[ g_{\mu \nu},\{\psi\}],
	\end{equation}
	where, once again, we have included a gauge fixing term $\lgr'_{\rm{gf}}$.	Most importantly, this term will not be the same as in standard gravity [cf.~Eq.~(\ref{gf})], since the scalar-tensor action does not have the same symmetries as the Einstein-Hilbert one, given that the non-minimal coupling affects the (non-null) geodesics. This requires us to upgrade the usual harmonic gauge condition.
	
	The first change is to replace the constant prefactor ${M_{\rm{Pl}}}^2$ by the non-minimal coupling, such that
	\begin{equation}\label{gf_mod1}
		-\frac{{M_{\rm{Pl}}}^2}{4}g_{\alpha \beta}\Gamma^\alpha\Gamma^\beta\to -\frac{F(X)}{4}g_{\alpha \beta}\Gamma^\alpha\Gamma^\beta.
	\end{equation}
	Second, we want to make sure that the gauge condition reflects the symmetries of the action, which are modified relative to the Einstein-Hilbert case\footnote{We might be tempted to work with only the first modification in Eq.~\eqref{gf_mod1} for the updated gauge fixing term. While doing so does not affect the non-relativistic limit for the fifth forces, this choice significantly complicates the calculations and leads to apparent deviations from the corresponding Einstein-frame results in the relativistic limit.}. Thus, the scalar-tensor equivalent of the harmonic condition must be defined via
	\begin{equation}\label{DDjf}
	    D^\mu D_\mu=\partial^\mu \partial_\mu,
	\end{equation}
	where $D^\mu$ is the covariant derivative as constructed with respect to the Jordan-frame action. This is the focus of the next subsection.
	
	
\subsubsection{Updating the covariant derivative}

	We know that General Relativity must be symmetric under diffeomorphisms. Thus, we will study which conditions the covariant derivative must satisfy in order to leave the action invariant under transformations of the form
	\begin{equation}\label{isometry}
		g^{\mu \nu}\to g^{\mu \nu} + D^\mu \xi^\nu +D^\nu \xi^\mu,	
	\end{equation}
	where $D_\mu$ is generically defined when acting on a four vector $Y^\nu$ as
	\begin{equation}
		D^\mu Y^\nu=	\partial^\mu Y^\nu + g^{\mu\rho}\Gamma_{\rho \sigma}^\nu Y^\sigma+ C^\mu Y^\nu,
		\label{newcov}
	\end{equation}
	and $C^\mu$ contains the possible contributions of $X$ to the isometries of the spacetime.
	
	To find whether the action is symmetric under diffeomorphisms, we first need to vary the gravitational part of the action [Eq.~\eqref{Action-BD-GF}] with respect to the metric, leading to
	\begin{equation}
		\delta S= -\int \D^4{x} \sqrt{-g} \frac{F(X)}{2} G_{\mu \nu} \delta g^{\mu\nu},
	\end{equation}
	where $G_{\mu \nu}$ is the Einstein tensor and we take the small variation of the metric $\delta g^{\mu\nu}$ to be given by the isometry from Eq.~\eqref{isometry}. Since we expect this transformation to be a symmetry of the action, $\delta S$ should vanish, i.e.,
	\begin{equation}
		\delta S= -\int \D^4{x} \sqrt{-g} F(X) G_{\mu \nu} D^{\left(\mu\right.} \xi^{\left. \nu\right)}=0,
	\end{equation}
	where the parentheses indicate the symmetrization of indices, such that
	\begin{equation}
	    D^{\left(\mu\right.} \xi^{\left. \nu\right)}=\frac{1}{2}(D^\mu\xi^\nu +D^\nu \xi^\mu).
	\end{equation}
	Inserting the covariant derivative from Eq.~\eqref{newcov} and integrating by parts, we find
	\begin{equation}
		\delta S= \int \D^4{x} \sqrt{-g} [\nabla^\nu (F(X)G_{\mu \nu}) - F(X) G_{\mu\nu}C^\nu ]\xi^\mu=0,
	\end{equation}
	where 
	\begin{equation}
		\nabla_\mu Y^\nu=	\partial_\mu Y^\nu + \Gamma_{\mu \sigma}^\nu Y^\sigma
	\end{equation}
	is the covariant derivative of standard gravity. Since both $\nabla^\nu$ and $G_{\mu\nu}$ only depend on the metric, they will still satisfy the Bianchi identity, leading to
	\begin{equation}
	    \nabla^\nu G_{\mu \nu}=0.
	\end{equation}
	However, as $\nabla^\nu$ also acts on $F(X)$, the perturbed action will take the following form
	\begin{equation}
		\delta S= \int \D^4x \sqrt{-g}[\nabla^\nu F(X) - F(X) C^\nu]G_{\mu\nu}\xi^\mu=0.
	\end{equation}
	Note that, for a constant $F(X)$, this equation would vanish, showing that the geodesics of the Einstein-Hilbert action are in agreement with those from standard gravity, encoded in $\nabla_\mu$. However, to have a vanishing perturbed action in the Jordan frame, we need
	\begin{equation}
		C^\nu=\frac{\partial^\nu F(X)}{F(X)}.
	\end{equation}
	Hence, we see that the non-minimal coupling of $X$ alters the isometries of the spacetime.
	
	The new contribution $C^\mu$ to the covariant derivative can be also expressed as a modification of the Christoffel symbols [i.e., as a $(1,2)$-form]. While this full expression is not needed to update the harmonic gauge condition, it is useful for explicit calculations of geodesics. While the derivation of the $(1,2)$-form can be done by studying the symmetries of the Jordan-frame action, one can instead (since we know that it exists) use the Einstein frame as a shortcut, where standard gravity is recovered. Since we already know the symmetries of the Einstein-frame action, we can fix the gauge in that frame and then undo the coordinate transformation. In this way, we will get the corresponding set of symmetries for the Jordan frame.  
	
	For instance, Weyl transforming the covariant derivative of the vector $Y_\nu$ from the Einstein to the Jordan frame, we find that
	\begin{equation}
	\tilde\nabla_\mu Y_\nu \to D_\mu Y_\nu=\partial_\mu Y_\nu - \Gamma^\rho _{\mu \nu} Y_\rho + C^\rho_{\mu \nu} Y_\rho,	
	\end{equation}
	where
	\begin{equation}
		C_{\mu \nu}^\sigma =-\frac{1}{2F(X)}\left[2\delta^\sigma_{\left(\mu\right.}\partial_{\nu\left.\right)}F(X) -g_{\mu \nu}\partial^\sigma F(X)\right].
	\end{equation}
	This is the correction to the Christoffel symbols coming from the breaking of the weak equivalence principle. Moreover, taking the trace of this term, we find that it is consistent with our previous result:
	\begin{equation}
		g^{\mu\nu} C_{\mu \nu}^\sigma=\frac{\partial^\sigma F(X)}{F(X)}=C^\sigma.
		\label{traceC}
	\end{equation}
	
	Thus, in contrast to standard gravity, where the geometry of spacetime is exclusively modified through the Christoffel symbols, the non-minimal coupling of the field $X$ alters the isometries, accounting for the breaking of the weak equivalence principle.


\subsubsection{Gauge fixing and obtaining the generic linearized Jordan-frame Lagrangian}
	
	Having derived the scalar-tensor covariant derivative, we can now continue upgrading the harmonic gauge from Eq.~\eqref{gf_mod1}. Returning to Eq.~\eqref{DDjf}, the harmonic condition imposes
	\begin{equation}
		D^\mu D_\mu=g^{\mu\nu}\left[\partial_\nu \partial_\mu-\Gamma ^\sigma_{\mu \nu}\partial_\sigma + C_{\mu \nu}^\sigma \partial_\sigma \right]=\partial^{\mu}\partial_{\mu}.
	\end{equation}
Thus, to satisfy the scalar-tensor harmonic gauge, it follows that
	\begin{equation}
		\Gamma^\mu - \frac{\partial^\mu F(X)}{F(X)}=0.
	\end{equation}
	Following the same logic as in Eq.~\eqref{HG-constraint}, we now replace each $\Gamma^\alpha$ from Eq.~\eqref{gf_mod1} by the new constraint. With this, we introduce the {\it scalar-harmonic gauge} via
	\begin{equation}
		\lgr'_{\rm{gf}}=-\frac{F(X)}{4}g_{\mu\nu}\left[ \Gamma^\mu -\frac{\partial^\mu F(X)}{F(X)}\right]\left[\Gamma ^\nu -\frac{\partial^\nu F(X)}{F(X)}\right].
		\label{gfjf1}	
	\end{equation}
	
	It is important to point out that this is not the first time this gauge has been used. For instance, it is mentioned by Fuji and Maeda~\cite{FM} and used in a number of papers where the authors employ the background covariant DeWitt condition \cite{DeWitt:1964mxt,Barvinsky:1985an,Steinwachs:2011zs}. However, here, we have presented a different way to find the gauge symmetries of the Jordan frame. Moreover, unlike in previous papers, we define the gauge fixing term using the complete metric, which allows us to perturb  consistently to higher orders in the fluctuations.
	
	Note also that we have been able to define this Jordan-frame gauge fixing term because we already had the standard gravity condition defined at the Lagrangian level [Eq.~\eqref{gf}]. This might be problematic when working with other gauges that are defined only after gravity has been linearized. For instance, the Newtonian gauge does not have a full metric expression; it is defined by setting certain modes of the graviton to zero. Therefore, we cannot apply our gauging method in this case, which could lead to misinterpreted results when going to higher orders in the perturbations, since additional terms should appear because of the new symmetries of the action.
	
	For the gauge fixing term in Eq.~\eqref{gfjf1}, it proves convenient to divide it into three parts:
\begin{itemize}
    \item [(i)] a contribution to the graviton kinetic energy:
    \begin{equation}
    \lgr_{{\rm{G}}}=-\frac{F(X)}{4}g_{\mu\nu}\Gamma^\mu\Gamma^\nu;
    \end{equation}
    
    \item [(ii)] a contribution to the kinetic mixing:
    \begin{equation}
        \lgr_{\rm{KM}}=\frac{1}{2}\Gamma^\mu\partial_\mu F(X) ;
    \end{equation}
    
    \item [(iii)] a contribution to the $X$ kinetic energy:
    \begin{equation}
        \lgr_{\rm{SF}}=-\frac{1}{4F(X)}g_{\mu \nu}\partial^\mu F(X) \partial^\nu F(X);
    \end{equation}
\end{itemize}
	such that 
	\begin{equation}
	\lgr'_{\rm{gf}}=\lgr_{\rm{G}}+\lgr_{\rm{KM}}+\lgr_{\rm{SF}}.
	\end{equation} 
	
	Inserting this gauge fixing term into Eq.~\eqref{Action-BD-GF} and linearizing the action similarly to the standard gravity case, we obtain the following expansion up to second order for the associated Lagrangian:
	\begin{align}
			\lgr=&\frac{F(X)}{4}\left[ \frac{1}{4}\partial_\mu h \partial^\mu h -\frac{1}{2}\partial_\rho h_{\mu \nu}\partial^\rho h^{\mu \nu}\right]
			+ \frac{F(X)}{2}\left[\partial^\mu\partial^\nu h_{\mu \nu} - \Box h\right]\nonumber\\& + \frac{F'(X)}{2}\left[\partial^\lambda h_{\lambda\mu} - \frac{1}{2}\partial_{\mu}h\right]\partial^\mu X
			-\frac{1}{2}\left[Z(X)+\frac{F'(X)}{2F(X)}^2\right]g^{\mu\nu}\partial_\mu X\partial_\nu X - U(X)\nonumber\\& + \frac{1}{2}h^{\mu\nu}T_{\mu\nu}+ \lgr_{\rm{m}}(\eta_{\mu\nu}).
	\end{align}
	The would-be total derivative term in the case of standard gravity, corresponding to the terms involving second derivatives of the metric fluctuations, no longer vanishes on integration by parts. Moreover, it creates an additional kinetic mixing term that, once added to $\lgr_{\rm KM}$, yields
	\begin{align} 
			\lgr=&\frac{F(X)}{4}\left[ \frac{1}{4}\partial_\mu h \partial^\mu h -\frac{1}{2}\partial_\rho h_{\mu \nu}\partial^\rho h^{\mu \nu}\right]
			+ \frac{F'(X)}{4}\eta^{\mu\nu}\partial_\mu h \partial_\nu X \nonumber\\
			-&\frac{1}{2}\left[Z(X)+\frac{F'(X)}{2F(X)}^2\right]\eta^{\mu\nu}\partial_\mu X\partial_\nu X - U(X) \label{4}
			+ \frac{1}{2}h^{\mu\nu}T_{\mu\nu}+ \lgr_{\rm{m}}\{\eta_{\mu\nu}\}.
	\end{align}
	This is the generic effective Lagrangian of the linearized theory in the Jordan frame. When the function $F(X)$ is constant, the expression reduces to that of General Relativity, as expected. Thus, the main difference is the new kinetic mixing term, and it is this which can lead to fifth forces between matter fields. 
	
	In contrast to the Einstein frame, the main contribution to the fifth-force coupling, as analysed in the Jordan frame, is via the kinetic mixing and not a mass mixing. This comes from the fact that the graviton propagator ($\propto 1/q^2$) cancels the momentum dependence of the mixing vertex ($\propto q^2$) in every oscillation between the field $X$ and the graviton, such that, unlike the case of a massive field, there is no additional momentum suppression in the non-relativistic limit. In the next section, we will calculate explicitly the fifth forces that arise through this kinetic mixing in the Jordan frame.
	
	
\section{Fifth forces in the Jordan frame}\label{fourth}
	
	 Having derived the general expression for the Lagrangian up to second order in the metric fluctuations for the Brans-Dicke-type scalar-tensor theories in Eq.~\eqref{4}, we now turn our attention to specific models to see how the fifth forces arise from the kinetic mixings between the graviton and the non-minimally coupled field $X$. We will show that the results agree with those obtained previously in the Einstein frame.
	
	The action corresponding to Eq.~\eqref{BDactionX}, with a matter sector given by Eqs.~\eqref{sm} and \eqref{comb_pot}, is
	\begin{align} \label{BDpureparam}
			S=\int \D^4{x} \sqrt{-g} \, &\bigg{[}\frac{X}{2}R +\lgr'_{\rm{gf}} -\frac{\omega(X)}{2X}g^{\mu \nu}\dmu X \dnu X - \frac{1}{2}g^{\mu \nu}\dmu \Phi \dnu \Phi  \nonumber\\ 
			-&	\frac{1}{2} g^{\mu \nu} \partial_{\mu} \Theta  \partial_{\nu} \Theta+\frac{1}{2} \mu_{\theta}^{2} \frac{X}{\Mt} \Theta^{2}-\frac{\lambda_{\theta}}{4 !} \Theta^{4}-\frac{3}{2} \frac{\mu_{\theta}^{4}}{\lambda_{\theta}} \frac{X^2}{\Mt^2}\, \nonumber\\
			-&\bar{\psi}i{\stackrel{\leftrightarrow}{\slashed{\partial}}}\psi - y\bar{\psi}\Phi\psi -U(\Phi,\Theta)\bigg{]},
	\end{align} 
	where $U$ is given by Eq.~\eqref{comb_pot}.  Note that we have already substituted in Eq.~\eqref{BDpureparam} for the Brans-Dicke model functions~\cite{Brans:1961sx}, with
	\begin{equation}
		F(X)=X,\qquad Z(X)=\frac{\omega(X)}{X}.
		\label{5}
	\end{equation}
	We assume that $\omega(X)$ is a slowly varying function that can be taken effectively constant. 
	
	We now proceed to linearize the Lagrangian, making use of the results from the preceding section. Substituting Eq.~\eqref{5} into Eq.~\eqref{4}, we thus find
	\begin{align}
			\lgr=&\frac{X}{4}\left[ \frac{1}{4}\partial_\mu h \partial^\mu h -\frac{1}{2}\partial_\rho h_{\mu \nu}\partial^\rho h^{\mu \nu}\right] - \frac{1}{2}\frac{2\omega+1}{2X}\eta^{\mu\nu}\partial_\mu X\partial_\nu X- \frac{1}{2}\eta^{\mu\nu}\partial_\mu \Phi\partial_\nu \Phi \nonumber \\
			+& \frac{1}{4}\eta^{\mu\nu}\partial_\mu h \partial_\nu X -U(\Phi,\Theta)
			- \frac{1}{2}\eta^{\mu\nu}\partial_\mu \Theta\partial_\nu\Theta + \frac{1}{2} \mu_{\theta}^{2} \frac{X}{\Mt} \Theta^{2}-\frac{\lambda_{\theta}}{4 !} \Theta^{4}-\frac{3}{2} \frac{\mu_{\theta}^{4}}{\lambda_{\theta}} \frac{X^2}{\Mt^2}\nonumber\\
			+& \frac{1}{2}h^{\mu\nu}T_{\mu\nu}-\bar{\psi}i{\stackrel{\leftrightarrow}{\slashed{\partial}}}\psi - y\bar{\psi}\Phi\psi+  \cdots.
	\end{align}
	where the ellipsis indicates terms higher than second order in $h_{\mu \nu}$.
	
	When linearizing around the background solution for $X$, namely $v_X$, it is possible to canonically normalize all the fields, including the graviton. To do so, we assume the background value of $X$ to vary very slowly compared to the other fields. This is a reasonable assumption, given that the tightest constraints from the analysis of the Moon's orbit~\cite{Muller:2007zzb} set the Planck mass to be almost constant at late times, a result that can be obtained by considering the impact of the late-time  Hubble friction on the evolution of $X$. Defining 
    \begin{equation}
	X=\frac{\chi^2}{2(2\omega + 1)}
		\label{mplsb}
	\end{equation}
	and making the replacement $h_{\mu\nu}\to2h_{\mu\nu}/M_{\rm{Pl}}$,
    	where
    	\begin{equation}\label{Eq:M_pl}
    	 	M^2_{\rm{Pl}}=\frac{v_\chi^2}{2(2\omega +1)},
    	\end{equation}
	the Lagrangian then takes the form
	\begin{align}
			\lgr=& \frac{1}{4}\partial_\mu h \partial^\mu h -\frac{1}{2}\partial_\rho h_{\mu \nu}\partial^\rho h^{\mu \nu} - \frac{1}{2}\eta^{\mu\nu}\partial_\mu \chi\partial_\nu \chi -  \frac{1}{2}\eta^{\mu\nu}\partial_\mu \Phi\partial_\nu \Phi+\frac{1}{\sqrt{2(2\omega + 1)}}\eta^{\mu\nu}\partial_\mu h \partial_\nu \chi \nonumber \\
			-& \frac{1}{2}\eta^{\mu\nu}\partial_\mu \Theta\partial_\nu \Theta+U(\Phi, \Theta)+\frac{1}{2} \mu_{\theta}^{2} \frac{\chi^2}{\tilde{M}'^2} \Theta^{2}-\frac{\lambda_{\theta}}{4 !} \Theta^{4}-\frac{3}{2} \frac{\mu_{\theta}^{4}}{\lambda_{\theta}} \frac{\chi^4}{\tilde{M}'^4}\label{laggrav}\nonumber\\
			+& \frac{1}{M_{\rm{Pl}}}h^{\mu\nu}T_{\mu\nu}-\bar{\psi}i{\stackrel{\leftrightarrow}{\slashed{\partial}}}\psi - y\bar{\psi}\Phi\psi+  \cdots,
	\end{align}
	where 
	\begin{equation}
	    \tilde{M}'^2=\Mt^2 2({2\omega +1}).
	\end{equation}
	Note that $M^2_{\rm{Pl}}$ is the effective gravitational coupling in the Jordan frame, while $\Mt^2$ is the one defined in the Einstein frame. Even though they belong to different frames, the conformal transformations appear to have forced them into the same Lagrangian. However, as we will see, this is not the case as the $\Mt^2$ dependence cancels out, leading to a final result in terms of $M^2_{\rm{Pl}}$, as expected. The next step is to diagonalize the mass matrix so that we can isolate the expected massless mode that can mediate any long-range fifth force.
    
	\subsection{Diagonalization}
	
    By considering the mass mixing terms from Eq.~(\ref{laggrav}), we can construct the following mass matrix:
	\begin{equation}
		m^2=
		\begin{pmatrix}
			m^2_\Phi & -Am_\Phi  & 0\\
			-Am_\Phi & m^2_\Theta & -Bm_\chi \\
			0 & -Bm_\chi & m^2_\chi
		\end{pmatrix},
	\label{massmat}
	\end{equation}
	where
	\begin{subequations}
	\begin{gather}
		m^2_\Phi=\frac{\lambda v_\Phi^2}{3},\qquad m_\Theta^2 =\frac{\beta^2}{2\lambda}v^2_\Theta + \frac{\mu^2_\theta}{ \Mp^2}v_\chi^2, \qquad m_\chi^2=\frac{\mu^2_\theta}{ \Mp^2}v_\Theta^2,\\
		v^2_\Phi=\frac{6\mu^2 +\beta{v_\Theta}^2}{\lambda}, \qquad v_\Theta^2 =\frac{3\mu^2_\theta }{\lambda_\theta \Mp^2}v_\chi^2, \qquad v_\chi^2=\frac{\lambda_\theta \Mp^2}{6\mu^2_\theta}v_\Theta^2,\\
		A^2=\frac{\beta^2}{2\lambda}v^2_\Theta,\qquad B^2= \frac{\mu^2_\theta}{ \Mp^2}v_\chi^2.
		\label{AB}
	\end{gather}
	\end{subequations}
	Diagonalizing this mass matrix, we obtain a new set of fields $ \phi$, $ \theta$ and $\sigma$, whose squared mass eigenvalues are
	\begin{equation}
		 m^2_{{\phi},{\theta}}=\frac{m^2_\Phi+m^2_\Theta+m^2_\chi\pm\sqrt{(-m^2_\Phi -A^2 +B^2 +m^2_\chi)^2 +4A^2B^2}}{2},\qquad	\label{mpt}
		 m_\sigma^2=0,
	\end{equation}
	wherein we see the anticipated massless mode $\sigma$.
	
	To determine how the original fields depend on these three modes, we need to find the eigenvectors of the mass matrix~(\ref{massmat}). After some algebra, we can show that
			\begin{align}	
			&\phi={N_{\phi}}
			\begin{pmatrix}
				\frac{\beta v_\Theta v_\Phi}{3(m^2_\Phi -C +D)} \\
				1 \\
				\frac{\mu^2_\theta v_\Theta v_\chi}{\Mp^2(m^2_\chi -C +D)}
			\end{pmatrix}&
			&\theta={N_{\theta}}
			\begin{pmatrix}
				\frac{\beta v_\Theta v_\Phi}{3(m^2_\Phi -C -D)} \\
				1 \\
				\frac{\mu^2_\theta v_\Theta v_\chi}{\Mp^2(m^2_\chi -C-D)}
			\end{pmatrix}&
			&	\sigma={N_\sigma}
			\begin{pmatrix}
				\frac{\beta v_\Theta}{\lambda v_\Phi} \\
				1 \\
				\frac{v_\chi}{v_\Theta}
			\end{pmatrix},&
		\label{evec}
		\end{align}
		where $N_{\phi}$, $N_{\theta}$ and $N_\sigma$ are normalization factors, and
		\begin{align}
			C=\frac{m^2_\Phi+m^2_\Theta+m^2_\chi}{2},\qquad D=\frac{\sqrt{(-m^2_\Phi -A^2 +B^2 +m^2_\chi)^2 +4A^2B^2}}{2}.
		\end{align} 
	
	For the fifth-force contribution to the M\o{}ller scattering, we need only expand the $\chi$ and $\Phi$ fields in terms of the massless eigenmode, since they are the only ones coupling to the fermion and graviton directly.
    The relevant expansions take the forms
	\begin{equation}
	  	\chi= \frac{a}{N_{{\phi}}}{\phi}+\frac{b}{N_{{\theta}}}{\theta}+\frac{c}{N_\sigma}\sigma,\qquad
	  	\Phi=\frac{a'}{N_{{\phi}}}{\phi}+\frac{b'}{N_{{\theta}}}{\theta}+\frac{c'}{N_\sigma}\sigma, \label{eq:field_exp}
	\end{equation}
	where $\{a,b,...\}$ are constant coefficients, which should not be confused with those appearing in Eq.~\eqref{para}. Since we are interested in the massless mode, we only need to determine $c$ and $c'$, and, after some algebra, we have
	\begin{subequations}
	\begin{align}
		c&=\frac{\theta_3-\phi_3}{(\theta_1 -\phi_1)(\sigma_3 - \phi_3) + (\sigma_1 -\phi_1)(\phi_3-\theta_3)},
		\label{c}\\
		c'&=-\frac{\theta_1-\phi_1}{(\theta_1 -\phi_1)(\sigma_3 - \phi_3) + (\sigma_1 -\phi_1)(\phi_3-\theta_3)},	
		\label{cp}
	\end{align}
	\end{subequations}
    where the subscripts refer to each component of the eigenvectors defined in Eq.~\eqref{evec}, without the corresponding normalizing factor $N_{\{\phi,\theta,\sigma\}}$.
    We are now in a position to derive an expression for the effective Lagrangian in terms of the massless mode and subsequently calculate its contribution to the M\o{}ller scattering.


\subsection{Non-relativistic fifth-force potential}

\begin{figure}
        \centering
        \includegraphics[scale=0.5]{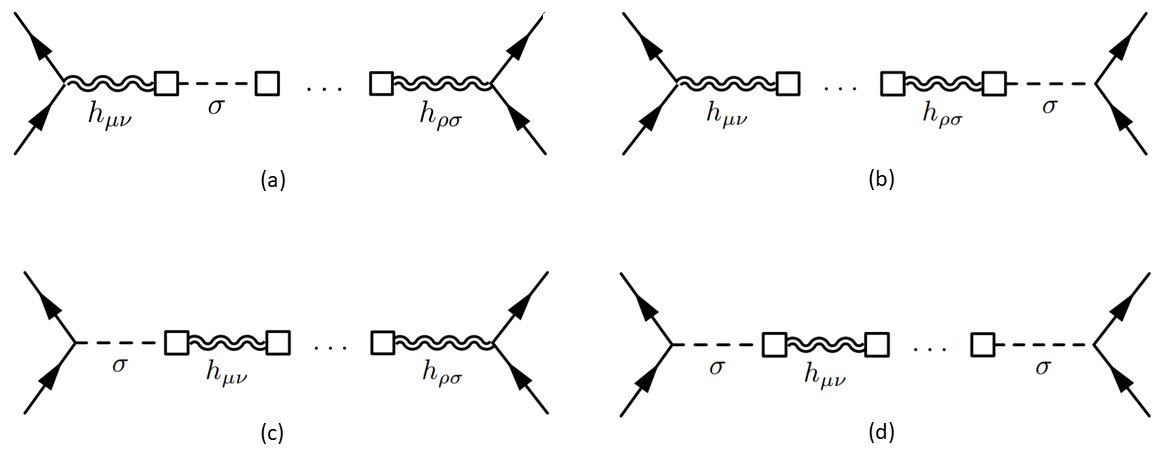}
        \caption{The diagrams that contribute to the M\o{}ller scattering in the Jordan frame. }
        \label{fig:sigmakinetic2}
    \end{figure}

	After diagonalizing the mass terms in the Lagrangian, we have found all the different ways that the long-range fifth forces can couple to the matter fields. From the linearization of scalar-tensor gravity, the fifth forces arise through the kinetic mixing between the graviton and the $\sigma$ field. In addition, after diagonalizing the mass terms, a new coupling between the massless mode and the fermion field appears as a result of their Yukawa interaction with the $\Phi$ field.  Thus, there are four distinct Feynman diagrams contributing to the M\o{}ller scattering, and these are shown in Fig.~\ref{fig:sigmakinetic2}.

\begin{figure}
\fbox{
    \begin{tabular}{c c}
        \begin{minipage}[b]{0.4\linewidth}
            $\bullet$ Graviton propagator \cite{Donoghue:2017pgk}
            \begin{equation}
                \begin{gathered}
                 \includegraphics[scale=0.28]{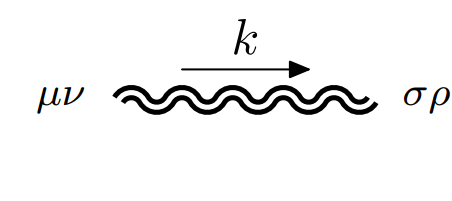}
                \end{gathered}
                = \frac{i P^{\mu \nu \sigma \rho}}{k^2}
                \nonumber
            \end{equation}
            \vspace{-7mm}
        	\begin{equation}
        		P^{\mu \nu \sigma \rho}=\frac{1}{2}\left(\eta^{\mu \sigma}\eta^{\nu \rho} + \eta^{\nu \sigma}\eta^{\mu \rho} -\eta^{\mu \nu}\eta^{\rho \sigma}\right)
        		\label{propp}
        		\nonumber
        	\end{equation}\\
        	\vspace{11mm}
        \end{minipage} & 
        \begin{minipage}[b]{0.4\linewidth}
            \vspace{8mm}
             $\bullet$ $\sigma$ field propagator
            \begin{equation}
                \begin{gathered}
                    \includegraphics[scale=0.28]{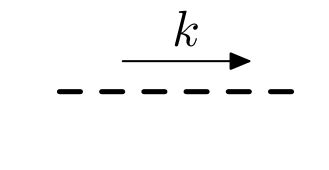}
                \end{gathered}
                =-\frac{i}{k^2}
                \nonumber
            \end{equation}
            \vspace{3mm}
           $\bullet$ Kinetic mixing
            \begin{equation}
                \begin{gathered}
                    \includegraphics[scale=0.22]{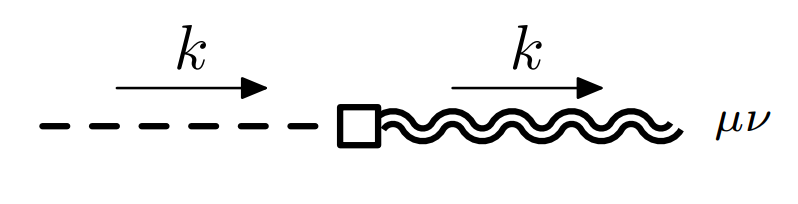}
                \end{gathered}
                =i \frac{\eta_{\mu \nu} k^2}{\sqrt{2(2\omega+1)}}
                \nonumber
            \end{equation}\\
        \end{minipage} \\
        \begin{minipage}{0.4\linewidth}
            \vspace{9mm}
            $\bullet$ Gravitational interaction \cite{Olyaei:2018asy}
            \begin{equation}
            	\begin{gathered}
                    \includegraphics[scale=0.21]{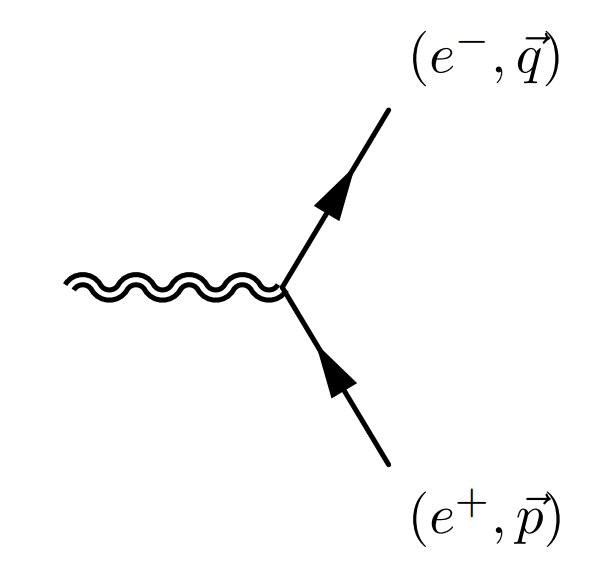}
                \end{gathered}
                = i \frac{\eta^{\mu\nu}\tau_{\mu \nu}}{M_{\rm{Pl}}}
        	    \nonumber
        	\end{equation}
        	\vspace{-5mm}
        	\begin{equation}
        		\tau_{\mu \nu}= 	\frac{1}{4}\left[(p+q)_\mu \gamma_\nu + \gamma_\mu (p+q)_\nu - 2\eta_{\mu\nu}\left(\slashed{q}+\slashed{p} -2m_e\right)\right]
        	    \nonumber
        	\end{equation}
        	\vspace{1mm}
        \end{minipage} &
        \begin{minipage}{0.4\linewidth}
             $\bullet$ Fermion-fermion-$\sigma$ interaction
             \vspace{-2mm}
            \begin{equation}
    \begin{gathered}
    \includegraphics[scale=0.21]{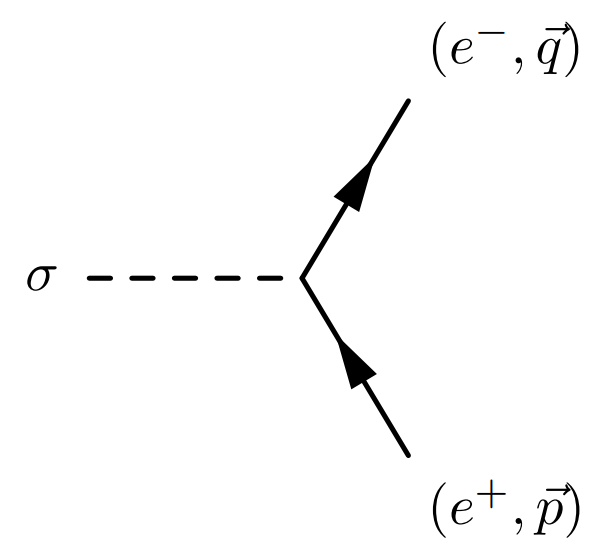}
    \end{gathered}
		= i \frac{yc'v_\chi}{cM_{\rm{Pl}}\sqrt{2(2\omega+3)}}\nonumber
    \end{equation}
    \vspace{5mm}
        \end{minipage}\\
    \end{tabular}\qquad\quad}
    \caption{Feynman rules for the Lagrangian [Eq.~\eqref{SHLag}] with an explicitly broken scale symmetry, where $\gamma_{\mu}$ are the gamma matrices. To a good approximation, we can take $cN^{-1}_\sigma\approx1$, since $\tilde{M}\gg1$. \label{fig:FeynExp}}
\end{figure}

    The terms in the Lagrangian relevant to the fifth force are as follows:
		\begin{align}
				\lgr_{\textrm{JF}}=&  \frac{1}{4}\partial_\mu h \partial^\mu h-\frac{1}{2}\partial_\rho h_{\mu \nu}\partial^\rho h^{\mu \nu}-\frac{1}{2}\eta^{\mu\nu}\partial_\mu \sigma\partial_\nu \sigma \nonumber \\
				+& \frac{cN^{-1}_\sigma}{\sqrt{2(2\omega + 1)}}\eta^{\mu\nu}\partial_\mu h \partial_\nu \sigma - yc'N^{-1}_\sigma \bar{\psi}\sigma\psi
				+ \frac{1}{M_{\rm{Pl}}}h^{\mu\nu}T_{\mu\nu} + \lgr_{\rm{m}}(\eta_{\mu\nu}),\label{SHLag}
		\end{align}
		and the resulting Feynman rules are summarized in Fig.~\ref{fig:FeynExp}.

   Since the structure of all the diagrams is very similar, we will describe only the contribution from Fig.~\ref{fig:sigmakinetic2}(a) in detail. The matrix element for this process is
	\begin{align}
			i\M_{\rm{\bf{(a)}}}&= \bar{u}(\mathbf{p}_1,s_1)\left( i\frac{\tau_{\mu \nu}}{{M}_{\rm{Pl}}}\right)u(\mathbf{p}_3,s_3)
			\left(i\frac{P^{\mu \nu a b}}{t}\right)\left(i \eta_{a b}\alpha_{\rm{K}} t\right)\left(\frac{i}{t}\right) \nonumber\\
			& \times \left[ \sum_{n=0}^{\infty} (i\alpha_{\rm{K}} t)^n\left(\frac{i\eta_{c d}P^{c d e f}\eta_{e f}}{t}\right)^n(i\alpha_{\rm{K}} t)^n \left(\frac{i}{t}\right)^n\right] 		\nonumber\\
			& \times (i\alpha_{\rm{K}} \eta_{g h}t)\left(\frac{iP^{g h \sigma \rho}}{t}\right) \bar{u}(\mathbf{p}_2,s_2)\left(i\frac{\tau_{\sigma \rho}}{{M}_{\rm{Pl}}}\right)u(\mathbf{p}_4,s_4),\label{amplitude_JF}
	\end{align}
	where, as before, $t=-(p_1 -p_3)^2$, $u(\mathbf{p},s)$ and $\bar{u}(\mathbf{p},s)$ are respectively the Dirac four-spinor and its Dirac conjugate, with spin projection $s$. Note that, for clarity, we have isolated each vertex and propagator with parentheses. For convenience, we have also defined the parameter
	\begin{equation}
	   \alpha_{\rm{K}}=\frac{1}{\sqrt{2(2\omega+1})}.
	\end{equation}
	Equation~(\ref{amplitude_JF}) can be simplified by making use of the following identities for $P^{\mu \nu \sigma \rho}$:
	\begin{align}
		\eta_{\mu \nu}P^{\mu \nu \sigma \rho}=-\eta^{\sigma \rho},\qquad \eta_{\mu \nu}P^{\mu \nu \sigma \rho}\eta_{\sigma \rho}=-4,
		\label{id}
	\end{align}
	and we find that we only have vertices involving the trace of $\tau_{\mu\nu}$, as we would have expected from Eq.~\eqref{eq:actionexpansion}.

Working in the non-relativistic limit and choosing the fermions to represent electrons with mass $m_e$, such that $p^{\mu}\sim q^{\mu}\approx (m_e,\vec{0})$, the spinors satisfy
	\begin{equation}
		\bar{u}(\mathbf{p},s)\gamma_\mu u(\mathbf{q},s')=2m_e \delta_{\mu 0}\delta_{s s'},
	\end{equation}
 	in which case, using the expression for $\tau=\eta^{\mu\nu}\tau_{\mu\nu}$ extracted from Fig.~\ref{fig:FeynExp}, we have
	\begin{equation}
		\bar{u}(\mathbf{p},s)\tau u(\mathbf{q},s')=-2m_e^2.
	\end{equation}
The matrix element then reduces to
	\begin{equation}
		\M_{\rm{\bf{(a)}}}=-\frac{1}{{M}^2_{\rm{Pl}}} \frac{4m^4_e\alpha^2_{\rm{K}}}{t}\left[\sum_{n=0}^{\infty} \left(-4\alpha_{\rm{K}}^2\frac{1}{t}\right)^n\right]\delta_{s_1 s_3}\delta_{s_2 s_4}		=-\frac{1}{{M}^2_{\rm{Pl}}} \frac{4m^4_e\alpha^2_{\rm{K}}}{(1+4\alpha^2_{\rm{K}})t}\delta_{s_1 s_3}\delta_{s_2 s_4}.
	\end{equation}

	To extract the non-relativistic potential, we take $t=-\mathbf{Q}^2$ (where  $\mathbf{Q}$ is the exchange momentum), and the contribution to the Yukawa potential is
	\begin{equation}
		V_{\rm{\bf{(a)}}}(r)=-\frac{1}{{M}^2_{\rm{Pl}}}\frac{m^2_e}{(4+\alpha^{-2}_{\rm{K}})}\int\frac{\D^3{\mathbf{Q}}}{(2\pi)^3}e^{i\mathbf{Q}\cdot\mathbf{x}} \frac{1}{\mathbf{Q}^2}
		=-\frac{1}{4\pi r}\frac{m^2_e}{{M}^2_{\rm{Pl}}2(2\omega +3)}.
		\label{Vcham}
	\end{equation}

The contributions from the remaining processes in Fig.~\ref{fig:sigmakinetic2} are
	\begin{subequations}
	\label{V_other}
	\begin{align}
	    V_{\rm{\bf{(b)}}}(r)=& V_{\rm{\bf{(c)}}}(r)=-\frac{1}{4\pi r}\left(\frac{c'v_\chi}{cv_\Phi}\right)\frac{m^2_e}{{M}^2_{\rm{Pl}}2(2\omega +3)},\\
	    V_{\rm{\bf{(d)}}}(r)=&-\frac{1}{4\pi r}\left(\frac{c'v_\chi}{cv_\Phi}\right)^2\frac{m^2_e}{{M}^2_{\rm{Pl}}2(2\omega +3)},
	\end{align}
	\end{subequations}
	and the sum of all the contributions to the Yukawa potential is
	\begin{equation}
	V_5(r)=-\frac{m_e^2}{4\pi rM_{\rm{Pl}}^2}\frac{\left(1+\frac{v_\chi \gamma}{v_\Phi}\right)^2}{2(2\omega+3)}.
	\label{Veffe}	
    \end{equation}
    After some algebra, we can show that
    \begin{equation}
        \label{eq:gammadef}
    	  \gamma=\frac{c'}{c}
    	  =-\frac{\beta v^2_\Theta}{\lambda v_\chi v_\Phi},
    	  \end{equation}
	and, using the fact that $v_\Theta^2=(\lambda v_\Phi^2-6\mu^2)/\beta$, we obtain the following final expression
    	\begin{equation}
    		V_5(r)=-\frac{1}{4\pi r}\frac{m_e^2}{M^2_{\rm{Pl}}2(2\omega +3)}\frac{4\mu^4}{m_\Phi^4},
    		\label{eq:V potential_spontaneous}
    	\end{equation}
    where we recall that
    \begin{equation}
    m_\Phi^2=2\mu^2+\frac{\beta v_\Theta^2}{3}.
    \end{equation}
    This is in perfect agreement with the result in the Einstein frame\footnote{Since conformal transformations modify the rulers used to measure distances, we must compare dimensionless quantities, which are unaffected by coordinate transformations. This could, e.g., be the ratio of the fifth-force potential to the standard Newtonian potential. This is to say that the expressions for the potentials should match but with $\Mt$ for the Einstein frame and $M_{\rm{Pl}}$ for the Jordan frame.} [Eq.~\eqref{solu}].
    Notice therefore that we also find that the fifth force vanishes in the absence of explicit scale breaking ($\mu\to0$), as we did in the Einstein frame.
    
	Before concluding this work, we consider in the next subsection the full expressions for the matrix element of the M\o{}ller scattering away from the non-relativistic limit.
	
	
	\subsection{M\o{}ller scattering for purely explicit scale breaking ($\beta \to 0$)}
	\label{calc:inconsistent}
	
    For simplicity, we will work in the purely explicit scale-breaking limit (i.e., $\beta \to 0$). We consider the spin-averaged squared matrix element
    	\begin{equation}
		\overline{|\mathcal{M}|^{2}}=\frac{1}{4}\sum_{ \text {spins }}|\mathcal{M}|^{2},
	\end{equation}
   
	After some algebra, we find the $t$-channel contribution
			\begin{align}
			\overline{|\mathcal{M}_{\rm{JF}}|^{2}}=&\frac{\tilde{M}^4}{M^4}\overline{|\mathcal{M}_{\rm{EF}}|^{2}}\nonumber\\=&\frac{1}{4{M}^4t^2} {\rm Tr} \left\lbrace \left[\frac{3}{2}(\slashed{p}_1+\slashed{p}_3)-4m_e\right](\slashed{p}_3+m_e)\left[\frac{3}{2}(\slashed{p}_1+\slashed{p}_3)-4m_e\right](\slashed{p}_1+m_e) \right\rbrace \nonumber\\
				&\times {\rm Tr} \left\lbrace \left[\frac{3}{2}(\slashed{p}_2+\slashed{p}_4)-4m_e\right](\slashed{p}_4+m_e)\left[\frac{3}{2}(\slashed{p}_2+\slashed{p}_4)-4m_e\right](\slashed{p}_2+m_e) \right\rbrace,
				\label{CS_SH}
		\end{align}
		showing the same frame covariance as in the Yukawa potential in the sense that the Jordan- and Einstein-frame results differ only by $M\to \tilde{M}$.\footnote{Had we not used the updated harmonic gauge condition in Eq.~\eqref{gfjf1}, we would have found agreement up to an additional numerical multiplicative factor.} The mass scales are related to the Planck masses of each theory through ${M}^2=2(2\omega +3) M_{\rm{Pl}}^2$.
	
	Including the $u$-channel contribution  ($u=-(p_4-p_1)^2$), and taking the ultra-relativistic limit (i.e., $m_e\to0$), we obtain:\footnote{To calculate the traces of the product of the gamma matrices, we used the Mathematica package FeynCalc~\cite{Shtabovenko:2020gxv,Shtabovenko:2016sxi,Mertig:1990an}.} 
	\begin{equation}
	    \overline{|\mathcal{M}_{\rm{JF}}|^{2}}=\frac{\tilde{M}^4}{M^4}\overline{|\mathcal{M}_{\rm{EF}}|^{2}}=\frac{81\left(t^2 + u^2 \right)}{16M^4}.
	\end{equation}
	We thus see that the scalar-harmonic gauge from Eq.~\eqref{gfjf1} leads to a perfect agreement between frames.


\section{Conclusion}\label{fifth}
	
	In this paper, we have studied the  fifth forces that can arise in scalar-tensor theories of gravity by considering the tree-level matrix elements directly in the Jordan frame. To do so, we had to perturb both the metric and the matter fields, requiring a consistent linearization of the modified gravity sector.
	
	We have shown that the fifth forces arise through a kinetic mixing between the graviton and the non-minimally coupled field in the Jordan frame. For the specific model described in this work, we have illustrated how the diagonalization of the mass matrix of the scalar sector yields an additional direct coupling between the massless fifth-force mediator and the fermion through the original Yukawa coupling of the would-be Higgs field. By this means, we were able to illustrate from the Jordan-frame perspective the role played by sources of explicit scale breaking in the matter sector in allowing fifth forces to couple to matter fields. 	
	
	In addition, we have shown that a full evaluation of the symmetries of the modified gravitational action is crucial for consistently updating the usual gauge fixing conditions. By imposing diffeomorphism invariance on the gravitational action in the knowledge that it can be mapped to Einstein gravity, we obtained a redefinition of the covariant derivative, wherein the breaking of the weak equivalence principle is manifest. In this way, we were led to an update of the usual harmonic gauge --- the so-called scalar-harmonic gauge [see Eq.~\eqref{gfjf1}] --- providing results in perfect agreement with those found in the Einstein frame. 
	
	While this gauge fixing term is not completely new, we have been able to define it at the level of the full metric. Up to second order, one can equivalently define the gauge using the DeWitt background condition~\cite{DeWitt:1964mxt, Barvinsky:1985an,Steinwachs:2011zs}. However, there are additional operators at order three and four in the graviton and additional scalar fields, which are captured by our full metric expression. 

This work forms a basis for considering the consistent gauging and performing perturbative analyses of fifth forces in other modified theories of gravity, including and especially those for which an Einstein frame does not exist. This may form the focus of future works.
	
\section*{Note added}
While preparing the arXiv preprint of this work for journal submission, a thesis was uploaded to the arXiv \cite{Riva:2021zli}, where similar techniques are used for the derivation of the scalar-harmonic gauge [Eq.\eqref{gfjf1}].
\begin{acknowledgements}
    The authors thank Clare Burrage for helpful discussions. This work was supported by the Science and Technology Facilities Council (STFC) Consolidated Grant [Grant No.\ ST/T000732/1] and STFC studentship [Grant No.\ ST/V506928/1]; a United Kingdom Research Innovation (UKRI) Future Leaders Fellowship [Grant No.\ MR/V021974/1]; a Nottingham Research Fellowship from the University of Nottingham; and a Leverhulme Research Fellowship [Grant No.\ RF-2021-312]. 
\end{acknowledgements}
	

\appendix


\section{Understanding the fifth-force couplings at the Lagrangian level}

In this appendix, we provide further details of how the theory from Eq.~\eqref{SHLag} behaves in the two scale-breaking limits: explicit and dynamical. To this end, we  diagonalize the kinetic terms by making the following transformations of the graviton and massless mode:
	\begin{equation}
		 h_{\mu \nu}\to h_{\mu \nu} +\frac{cN^{-1}_\sigma}{\sqrt{4cN^{-1}_\sigma + 2(2\omega+1)}}\sigma\eta_{\mu\nu},\qquad \sigma\to-\frac{\sqrt{2(2\omega +1)}}{\sqrt{4cN^{-1}_\sigma + 2(2\omega+1)}}\sigma.
	\end{equation}
	This leads to the Lagrangian
    	\begin{align} 
    		\lgr=& \frac{1}{4}\partial_\mu h \partial^\mu h-\frac{1}{2}\partial_\rho h_{\mu \nu}\partial^\rho h^{\mu \nu}  - \frac{1}{2}\eta^{\mu\nu}\partial_\mu \sigma\partial_\nu \sigma \nonumber\\
    		-&\frac{1}{M_{\rm{Pl}}}\frac{cN^{-1}_\sigma}{\sqrt{4cN^{-1}_\sigma + 2(2\omega+1)}}\sigma T + y\frac{c'N^{-1}_\sigma\sqrt{2(2\omega +1)}}{\sqrt{4cN^{-1}_\sigma + 2(2\omega+1)}}\bar{\psi}\sigma\psi
    		\label{lgflag}\nonumber\\
    		+& \frac{1}{M_{\rm{Pl}}}h^{\mu\nu}T_{\mu\nu}+ \lgr_\psi(\eta_{\mu\nu}) +\dots .
    		\end{align}
    Herein, we see that the massless mode couples both to the trace of the matter energy-momentum tensor and directly to the fermion field through a Yukawa coupling. We can now consider the two limits:

    \paragraph*{\bf Explicitly broken scale symmetry ($\boldsymbol{\beta\to0}$):}
	As we can see, the new Yukawa coupling depends linearly on $c'$, which, from Eq.~\eqref{cp}, is proportional to $\theta_1-\phi_1$. Extracting each term from Eq.~\eqref{evec}, we obtain
    \begin{equation}
    	\phi_1=\frac{\beta v_\Theta v_\Phi}{3(m^2_\Phi -C +D)},\qquad	\theta_1=\frac{\beta v_\Theta v_\Phi}{3(m^2_\Phi -C -D)},	
    \end{equation}
    such that there is no overlap between $\Phi$ and $\sigma$ in the limit $\beta\to 0$ and $c'$ vanishes. It follows that the Yukawa coupling of $\sigma$ arises through the dynamical scale breaking. In addition, we can show that
	\begin{equation}
	\label{eq:cNm1}
	cN^{-1}_\sigma\approx1,
	\end{equation}
	such that
	\begin{equation}
	\lgr\supset\frac{1}{M_{\rm{Pl}}\sqrt{2(2\omega+3)}}\sigma T,
	\end{equation}
	from which we readily recover the result for the Yukawa potential of the Brans-Dicke-type model [Eq.~\eqref{exp_eins}] in the purely explicit scale-breaking limit.
	
	\paragraph*{\bf Dynamically broken scale symmetry ($\boldsymbol{\mu\to0}$):}
	In this limit, neither of the interaction terms vanish. Instead, as we will now show, they exactly cancel against one another. In the non-relativistic limit, the trace of the energy-momentum tensor is
	  \begin{equation}
	  	T\approx m_e\bar{\psi}\psi, \qquad m_e=yv_{\Phi},
	  \end{equation}
	  where we have chosen the fermions to represent electrons of mass $m_e$. The two interaction terms can therefore be combined such that
     \begin{equation}
     \lgr\supset -\frac{m_ecN^{-1}_\sigma}{M_{\rm{Pl}}\sqrt{4cN^{-1}_\sigma +2(2\omega +1)}}\left(1+\frac{v_\chi \gamma}{v_\Phi}\right)\bar{\psi}\sigma\psi,	
      	\label{inter}
      \end{equation}
    where $M_{\rm Pl}$ is defined in Eq.~\eqref{Eq:M_pl}. Using the result for $\gamma$ from Eq.~\eqref{eq:gammadef}, we have
      \begin{equation}
      \lgr\supset -\frac{m_ecN^{-1}_\sigma}{M_{\rm{Pl}}\sqrt{4cN^{-1}_\sigma +2(2\omega +1)}}\left(1 - \frac{v_\chi}{v_\Phi}\frac{\beta v^2_\Theta}{\lambda v_\chi v_\Phi} 	\right)\bar{\psi}\sigma\psi.
      \label{eq:cancelmid}
      \end{equation}
    Subsequently recalling that 
      \begin{align}
      	v^2_\Theta=\frac{\lambda v_\Phi^2 }{\beta} - \frac{6\mu^2}{\beta},
      \end{align}
    we can rewrite the large parenthesis in Eq.~\eqref{eq:cancelmid} to give
      \begin{equation}
      \lgr\supset -\frac{m_ecN^{-1}_\sigma}{M_{\rm{Pl}}\sqrt{4cN^{-1}_\sigma +2(2\omega +1)}}\left( 1 -1 + \frac{6\mu^2}{\lambda v_\Phi^2}\right)\bar{\psi}\sigma\psi.
      \end{equation}
    Since $v_{\Phi}$ remains finite in the limit $\mu\to 0$, we indeed see that the two interaction terms exactly cancel in the limit of purely dynamical scale-symmetry breaking, such that the fifth force mediator $\sigma$ decouples from the matter fields, as expected (see, e.g., Refs.~\cite{Ferreira:2016kxi, Burrage:2018dvt}).

\end{document}